\documentclass{aa}   
\usepackage[utf8]{inputenc}
\usepackage[T1]{fontenc}
\usepackage{graphicx}
\usepackage{fancyhdr}
\usepackage{nomencl}
\usepackage{setspace}
\usepackage{enumitem}
\usepackage{etoolbox}
\usepackage{amsmath}
\usepackage{amssymb}
\usepackage{physics}
\usepackage{mathrsfs}
\usepackage{natbib}
\usepackage[colorlinks]{hyperref}
\usepackage{algpseudocode}
\usepackage{algorithm}
\usepackage{booktabs}
\usepackage{subfigure}
\usepackage{dsfont}
\usepackage{xcolor}
\usepackage{stmaryrd}
\usepackage{diagbox}
\usepackage{mathtools}
\usepackage{mdframed}
\usepackage{bm}
\usepackage{blindtext}
\usepackage[export]{adjustbox} 
\usepackage{multirow}
\usepackage{makecell}

\graphicspath{{./Images/}} 

\newcommand{\given}[1][]{\,#1\lvert\,}
\newcommand*{\tran}{^{\mkern-1.5mu\mathsf{T}}}

\definecolor{myblue}{RGB}{12, 12, 158}
\definecolor{myred}{RGB}{158, 19, 22}
\definecolor{myorange}{RGB}{245, 150, 12}
\definecolor{mygreen}{RGB}{26, 148, 49}
\definecolor{Prune}{RGB}{99,0,60}
\definecolor{Purple}{RGB}{75, 0, 130}
\definecolor{Pink}{RGB}{255, 105, 180}

\definecolor{deepskyblue}{RGB}{0, 191,255}
\definecolor{limegreen}{RGB}{50, 205, 50}

\hypersetup{
    citecolor=[rgb]{0.04, 0.20, 0.58},  
    linkcolor=[rgb]{0.86, 0.078, 0.235},
    urlcolor=[rgb]{0,0,1}
}

\definecolor{void}{RGB}{63,96,174}
\definecolor{wall}{RGB}{109,179,136}
\definecolor{filament}{RGB}{231,133,50}
\definecolor{node}{RGB}{217,33,32}
\definecolor{combination}{RGB}{83,158,182}

\bibpunct{(}{)}{;}{a}{}{,} 

\makeatletter
\renewcommand*\aa@pageof{, page \thepage{} of \pageref*{LastPage}}

\begin{document}

\title{Cosmology with cosmic web environments}
\subtitle{I. Real-space power spectra}

\author{Tony Bonnaire\inst{1,2, 3}, Nabila Aghanim\inst{1}, Joseph Kuruvilla\inst{1}, Aurélien Decelle\inst{2,4}}
\authorrunning{T. Bonnaire et al.}
\titlerunning{Cosmology with cosmic web environments I.}
\institute{
            Université Paris-Saclay, CNRS, Institut d'Astrophysique Spatiale, 91405, Orsay, France \\
            E-mail: \href{mailto:tony.bonnaire@universite-paris-saclay.fr}{tony.bonnaire@universite-paris-saclay.fr}
         \and
            Université Paris-Saclay, TAU team INRIA Saclay, CNRS, Laboratoire Interdisciplinaire des Sciences du Numérique, 91190, Gif-sur-Yvette, France
        \and
            Laboratoire de Physique de l’École normale supérieure, ENS, Université PSL,
            CNRS, Sorbonne Université, Université de Paris --- F-75005 Paris, France
        \and
            Departamento de Física Téorica I, Universidad Complutense, 28040 Madrid, Spain
        }
\abstract{
We undertake the first comprehensive and quantitative real-space analysis of the cosmological information content in the environments of the cosmic web (voids, filaments, walls, and nodes) up to non-linear scales, $k = 0.5$ $h$/Mpc. Relying on the large set of $N$-body simulations from the Quijote suite, the environments are defined through the eigenvalues of the tidal tensor and the Fisher formalism is used to assess the constraining power of the power spectra derived in each of the four environments and their combination. Our results show that there is more information available in the environment-dependent power spectra, both individually and when combined all together, than in the matter power spectrum. By breaking some key degeneracies between parameters of the cosmological model such as $M_\nu$--$\sigma_\mathrm{8}$ or $\Omega_\mathrm{m}$--$\sigma_8$, the power spectra computed in identified environments improve the constraints on cosmological parameters by factors $\sim 15$ for the summed neutrino mass $M_\nu$ and $\sim 8$ for the matter density $\Omega_\mathrm{m}$ over those derived from the matter power spectrum. We show that these tighter constraints are obtained for a wide range of the maximum scale, from $k_\mathrm{max} = 0.1$ $h$/Mpc to highly non-linear regimes with $k_\mathrm{max} = 0.5$ $h$/Mpc. We also report an eight times higher value of the signal-to-noise ratio for the combination of spectra compared to the matter one.
Importantly, we show that all the presented results are robust to variations of the parameters defining the environments hence suggesting a robustness to the definition we chose to define them.
}


\maketitle


\section{Introduction} \label{sect:introduction}

Among all the successes of modern cosmology is the observation, in both data and simulations, of the matter distribution at the megaparsec scales. This impressive pattern, commonly called the ``cosmic web'' \citep{Klypin1983, Bond1996}, was observed in first galaxy surveys like the Center for Astrophysics Redshift Survey \citep{deLapparent1986} and recently traced more precisely by other redshift surveys like the Sloan Digital Sky Survey \citep[SDSS,][]{York2000} or the Two-Micro All Sky Survey \citep[2MASS,][]{Skrutskie2006}.
In this multi-scale pattern, isolated clumps of matter residing in empty underdense parts of the Universe are flowing into flattened planar-like regions and are then funnelled into tubular elongated structures to finally end their journey feeding large and massive anchors.
The formation of these cosmic structures, respectively called voids, walls, filaments, and nodes, has been predicted by pioneering analytical models in the seventies \citep{Zeldovich1970, Shandarin1978} linking the gravitational collapse of the primordial density fluctuations, assumed and observed to be Gaussian distributed, to the formation of anisotropic structures.
This non-linear gravitational evolution yields a non-Gaussian distribution of the matter at late-time with strong mode couplings in which the information is spread over higher-order correlations. It is well-established that the way matter evolves and clusters through time and under the effect of gravity is highly impacted by the initial density perturbations and by the underlying cosmological model described by the values of the cosmological parameters.

Despite the non-Gaussian nature of the matter distribution at low redshifts, the two-point correlation function, or Fourier-equivalent power spectrum, constitutes a wealthy source of information about the cosmological model used in practice through the sparse and biased observations of tracers like galaxies to constrain cosmological parameters \citep[e.g.][]{Cole2005, Tinker2012, Alam2016, Gil-Marin2017}. Numerous works also show that the first higher-order term, namely the bispectrum, carries non-negligible information able to improve the constraints \citep[][]{Sefusatti2006, Yankelevich2019, Hahn2020, Hahn2021, Agarwal2021, Gualdi2021}. However, because of the difficulties of directly measuring and computing higher-order statistics \citep[see e.g.][]{Schmittfull2013, Philcox2021}, alternative methods were proposed to account only partially, or indirectly, for higher-order terms.
This is for instance the case of the marked statistics \citep{Stoyan1984} proposing a weighed version of the two-point information in which a mark is assigned to each tracer, i.e. galaxy, halo or particle, based for instance on the local luminosity \citep{Beisbart2000, Sheth2005} or density \citep{White2016}. More recently, the wavelet scattering transform \citep{Mallat2012} also introduced a non-linear transformation of the input density field to extract further information by cascading convolutions with directional wavelet filters and was successfully applied to cosmological simulations \citep[see e.g.][]{Allys2019, Allys2020, Cheng2020, Cheng2021}. These two approaches were shown to improve the constraints over the real-space matter power spectrum in \cite{Massara2021} and \cite{Valogiannis2021} respectively. Other works make use of direct additional observables to the density field, such as the velocity, to improve the constraints on cosmological models and parameters \citep[e.g.][]{Mueller2015, Kuruvilla2021, Kuruvilla2021a}.

While substantial efforts are made to link the properties of cosmic web elements (mainly clusters and filaments) to the formation and evolution of matter tracers like galaxies \citep[see e.g.][]{Alpaslan2014, Malavasi2017, Malavasi2021, Codis2018, Bonjean2017, Bonjean2020}, only little is currently known about the information these environments carry about the underlying cosmological model. A strong focus was placed on the use of the densest, hence more easily identified, elements of the cosmic web that are nodes. The hierarchical structure formation makes them particularly interesting for probing not only the matter and dark energy contents of the Universe \citep[e.g.][]{Bahcall1997, Bahcall1998, Holder2001, Salvati2018, Marulli2018} but also the amplitude of the initial density fluctuations \citep[for a review, see][]{Allen2011}.
The clustering properties of the other extreme environment represented by voids also drew the attention of cosmologists to probe the accelerated expansion of the Universe and the summed neutrino mass \citep{Lee2009, Lavaux2012, Pisani2015}.
The measure of voids' clustering, sizes, shapes, counts, bias and corresponding evolution with redshift are hence key quantities probing the underlying cosmological model \citep{Weygaert2011, Hamaus2014, Hamaus2015, Massara2015, Schuster2019, Kreisch2019}. The constraints brought by these two extreme environments that are voids and nodes are for instance combined in \cite{Bayer2021} and \cite{Kreisch2021} showing that the information provided by the halo mass function and the void size function leads to considerable improvement over the matter power spectrum constraints in real space.

The cosmic web complexity and full pattern, however, goes beyond the picture provided by the properties of voids and nodes. We can expect that, when splitting the matter into the several environments, a simple two-point statistics would deliver different information that, when combined, may break degeneracies and improve the constraints on the cosmological parameters. These expectations are in line with the recent findings of \cite{Paillas2021} in which the two-point correlation function computed from galaxies in several density bins improves the cosmological constraints in redshift-space. In this paper, we carry out the first thorough analysis of the cosmological information of the matter distribution in the several cosmic web environments identified by means of the eigenvalues of the tidal tensor and through the power spectrum statistic in the derived density fields. From the intrinsic differences in densities, tidal force and anisotropies these environments exhibit, we explore their distinct sensitivities to the variations of cosmological parameters.
After introducing the $N$-body simulations used in our analysis in Sect. \ref{sect:data}, we establish the methodological approach in Sect. \ref{sect:methodology} by presenting and assessing the environments definition together with defining the power spectrum estimator. Section \ref{sect:analysis} then presents the Fisher formalism and reports the constraints obtained on the six studied cosmological parameters using the power spectrum derived from each environment individually and from their combination. In Sect. \ref{sect:discussion}, we discuss the obtained results and compare them with other recent findings. After drawing some conclusions and perspective of this work in Sect. \ref{sect:conclusions}, we study the convergence and assumptions of the analysis in the appendices.

    \begin{figure}
        \centering
        \includegraphics[width=1\linewidth]{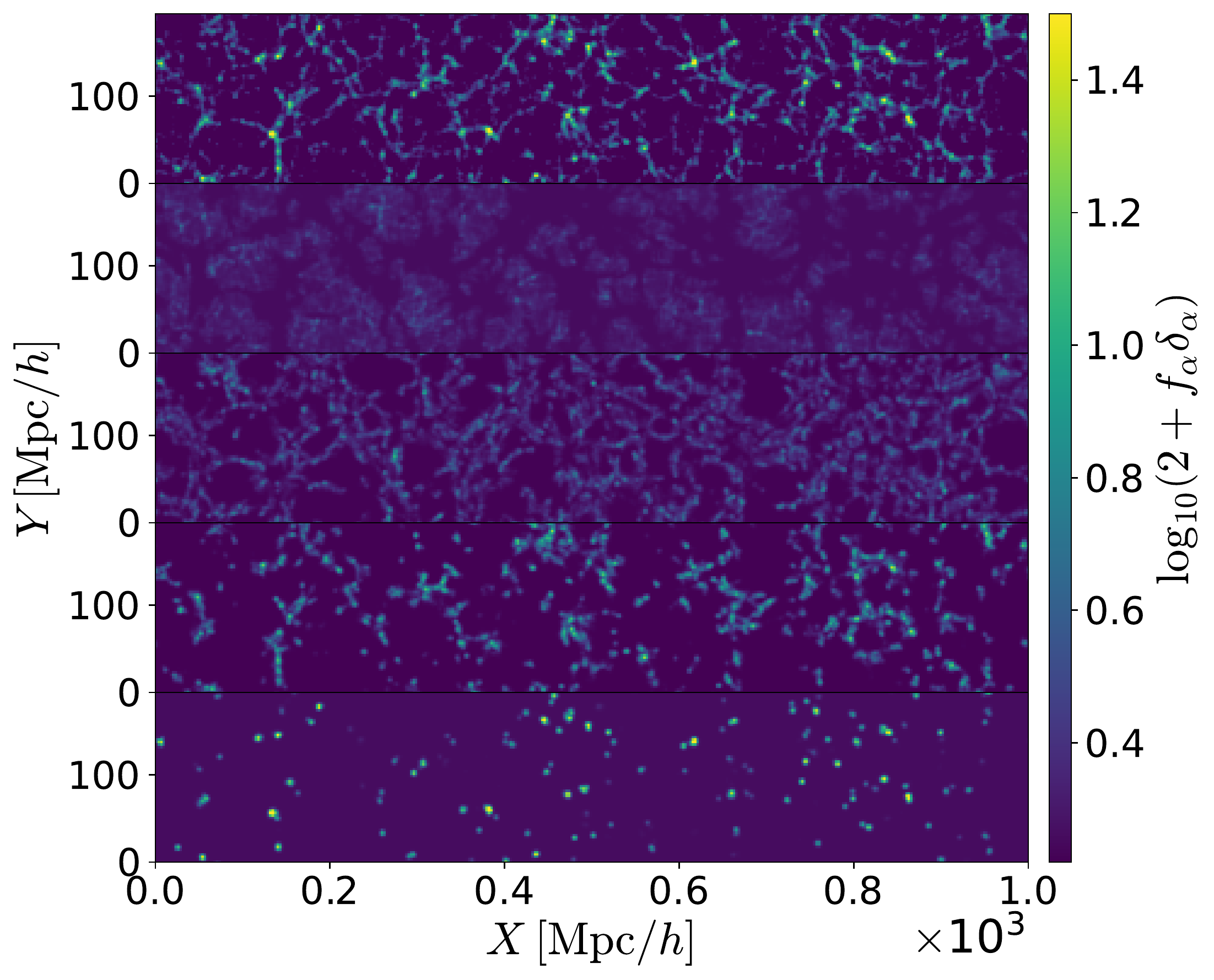}
        \caption{From top to bottom are shown a set of five $2.77$ Mpc/$h$ depth slices showing the fields $\delta_\mathrm{m}$, $\delta_\mathrm{v}, \delta_\mathrm{f}, \delta_\mathrm{w}$, and $\delta_\mathrm{n}$, respectively. Overdensity fields in cosmic web environments are computed from the T-web classification of particles and are normalised such that $\delta_\mathrm{m} = \sum_{\alpha} f_\alpha \delta_\alpha$.}
        \label{fig:fields_tweb}
    \end{figure}

\section{The Quijote simulations} \label{sect:data}

Quijote \citep{Villaescusa-Navarro2019} is a publicly available\footnote{\url{https://quijote-simulations.readthedocs.io/en/latest/}} large suite of $N$-body simulations. With $44\,100$ simulations spanning more than a thousand cosmological models, each with multiple realisations, it is the ideal dataset to perform statistical cosmological analyses as it allows to build accurate covariance matrices and compute derivatives for any cosmological representation. Each simulation consists of a set of $512^3$ particles (and $512^3$ neutrinos in massive neutrinos cases) that are evolved forward in time from $z=127$ to $z=0$ using a tree-PM Gadget-3 code \citep{Springel2005a} in a $L=1$ Gpc/$h$ size box initialised with the second-order Lagrangian perturbation theory for massless neutrino simulations and with the Zel'dovich approximation for massive neutrino ones.
The fiducial cosmology is a flat $\Lambda$CDM cosmology with parameters consistent with \cite{Planck2018}: $\Omega_\mathrm{m} = 0.3175$, $\Omega_\mathrm{b} = 0.049$, $h = 0.6711$, $n_\mathrm{s} = 0.9624$ and $\sigma_\mathrm{8} = 0.834$.
With these parameters, and assuming a zero mass for neutrinos ($M_\nu = 0$), $15,000$ random realisations are computed. The Quijote suite then provides $500$ realisations by varying individually each parameter, fixing the others at their fiducial values. The stepsizes are: $\dd \Omega_\mathrm{m} = 0.010$, $\dd \Omega_\mathrm{b} = 0.002$, $\dd h = 0.020$, $\dd n_\mathrm{s} = 0.020$, and $\dd \sigma_8 = 0.015$. Additionally, $500$ realisations using several sum of neutrinos mass are also computed, with $M_\nu = \sum m_\nu = \{0.1, 0.2, 0.4\}$ eV, that we will refer to as $M_\nu^{+}, M_\nu^{++}$, and $M_\nu^{+++}$ cosmologies respectively.
Because massive neutrino simulations are initialised with Zel'dovich approximation, we also use, for consistency, a massless neutrino simulation initialised this way for the computation of derivatives with respect to $M_\nu$.


\section{Cosmic web environments: segmentation and statistics} \label{sect:methodology}

    \subsection{Cosmic web segmentation}
    
        During the past decades, several methods were proposed to identify cosmic structures in either simulations or observations. Some algorithms directly rely on the density field as an input and base their definition on the curvature of the density or the gravitational field \citep[such as][]{Hahn2007, Forero-Romero2009, AragonCalvo2010, Nexus} or on its topological description \citep[][]{Spineweb, DisperseTheory}. Other algorithms are relying on more observable-like inputs such as the sparse distribution of galaxies (or halos in simulations) to identify structures such as filaments through geometrical principles \citep[][]{Stoica2007, Chen2015, Tempel2016} or based on graph theory \citep[such as][]{Barrow1985, Alpaslan2014, Bonnaire2020, Bonnaire2021b, Pereyra2020a}. In this theoretical work, we aim at studying all the cosmic web environments (voids, walls, filaments, nodes) based on the dark matter particles of the Quijote simulations.
        To achieve that, we choose a web finder that defines environments in a physical way based on the local level of tidal anisotropy using prescriptions originating from the linear growth of perturbations in the Zel'dovich approximation \citep{Zeldovich1970}. Following the formalism introduced in \cite{Hahn2007}, and later extended by \cite{Forero-Romero2009}, we thus identify the environments in the simulations through the tidal tensor $\bm{T}$, hereafter called T-web formalism.
        From the discrete set of particle positions, we first rely on a $B$-spline interpolation scheme \citep{Hockney1981, Sefusatti2016} to estimate the density field $\rho(\bm{x})$ on an $N_\mathrm{g}^3$ regular grid. For our purpose, we adopt an interpolation at the order four, namely the piecewise-cubic spline (PCS) scheme, in which the mass of a particle is spread over the $4^3 = 64$ closest cells. By noting $d=N_\mathrm{g}\lVert \bm{x} - \bm{x}_\mathrm{p} \rVert_2/L$ with $\bm{x}$ the centre of a grid cell, $\bm{x}_\mathrm{p}$ the particle position and $L$ the size of the box length (assuming a cubic box), PCS weights are given by
        \begin{equation}
        \begin{cases}
            (4 - 6d^2 + 3d^3)/6 & \mathrm{if} \, d \in \left[0, 1\right[, \\
            (2 - d)^3/6 & \mathrm{if} \, d \in \left[1, 2\right[, \\
            0 & \mathrm{otherwise}.
        \end{cases}
        \end{equation}
        This choice of interpolation order is a good trade-off between the accuracy of the reconstructed field and its computational time.
        
        From the density field $\rho$, we derive the gravitational potential $\Phi$ by solving the Poisson equation
        \begin{equation} \label{eq:poisson}
            \Delta \Phi(\bm{x}) = 4 \pi G \rho(\bm{x}),
        \end{equation}
        where $\Delta$ is the Laplacian operator and $G$ is the gravitational constant. It is convenient to write this equation in terms of the reduced gravitational potential $\Phi_r(\bm{x}) = \Phi(\bm{x}) / 4\pi G \bar{\rho}$, with $\bar{\rho}$ the averaged density, so that Eq.~\eqref{eq:poisson} satisfies $\Delta \Phi_r(\bm{x}) = \delta(\bm{x})$, with $\delta(\bm{x}) = \rho(\bm{x})/\bar{\rho} - 1$ the overdensity.
        Solving this reduced version of the Poisson equation in Fourier space using a discrete approximation of the Laplacian operator (in our case, a 7-point approximation) holds an estimate of $\Phi(\bm{x})$ on the grid. From the gravitational potential, we obtain the tidal tensor in each grid cell $\bm{x}$ as
        \begin{equation}
            \bm{T}_{i,j}(\bm{x}) = \frac{\partial^2 \Phi(\bm{x})}{\partial \bm{x}_i \partial \bm{x}_j},
        \end{equation}
        leading to the field of eigenvalues $\lambda_1(\bm{x}) \leq \lambda_2(\bm{x}) \leq \lambda_3(\bm{x})$.
        The cosmic environment associated with a grid cell $\bm{x}$ is finally obtained depending on the number of eigenvalues below a given threshold $\lambda_\mathrm{th}$, as defined in Table \ref{tab:tweb}. 
        
        We then use the segmentation of the density field obtained at the cell level to build individual overdensity fields for each environment. To do so, we simply propagate the classes at the particle level by assigning the same environment signature to all hosted particles in a given cell. Even though more sophisticated schemes have been proposed \cite[see e.g.][]{Wang2020}, we expect this approach to be sufficiently robust for the particles at hand, but could be of use when applying the procedure to coarser structures like halos. Based on the PCS interpolation scheme, we then build four corresponding overdensity fields $\{\delta_\mathrm{v}, \delta_\mathrm{w}, \delta_\mathrm{f}, \delta_\mathrm{n}\}$ where the subscripts respectively refer to void, wall, filament and node environment. The matter density field $\delta_\mathrm{m}$ is hence decomposed into the four environmental fields and fulfils the linear combination
        \begin{equation}
            \delta_\mathrm{m} = f_\mathrm{v} \, \delta_\mathrm{v} + f_\mathrm{w} \, \delta_\mathrm{w} + f_\mathrm{f} \, \delta_\mathrm{f} + f_\mathrm{n} \, \delta_\mathrm{n},
        \end{equation}
        where $f_\alpha$ denotes the mass fraction of the environment $\alpha$, namely $N_\alpha/N$ the number of particles in the $\alpha$ environment and $N$ the total number of particles\footnote{The mass fractions are here expressed in terms of number of particles since they all have the same mass in the $N$-body simulations.}.
        In Fig.~\ref{fig:fields_tweb} we show this decomposition with the contribution of each environment to the overall matter density field for a thin $2.77$ Mpc/$h$ depth slice. As expected, nodes describe a discrete set of dense objects found at the intersections of filaments and voids cover most of the surface with large low-density areas.
        
        \begin{table}
        \centering
        \caption{Cosmic web classification rules in the cell $\bm{x}$ depending on the eigenvalues $\lambda_1 \leq \lambda_2 \leq \lambda_3$ of the tidal tensor.}
        \label{tab:tweb}
        
        \smallskip
        
        \begin{tabular}{cl}
            \toprule
            Environment & Condition \\
            \midrule
            Void & $\lambda_1, \lambda_2, \lambda_3 < \lambda_\mathrm{th}$ \\
            Wall & $\lambda_1, \lambda_2 < \lambda_\mathrm{th}, \lambda_3 > \lambda_\mathrm{th}$ \\
            Filament & $\lambda_1 < \lambda_\mathrm{th}, \lambda_2, \lambda_3 > \lambda_\mathrm{th}$ \\
            Node & $\lambda_1, \lambda_2, \lambda_3 > \lambda_\mathrm{th}$ \\
            \bottomrule
        \end{tabular}
        \end{table}

        \subsection{Choice of the parameters}
        
         In our implementation of the T-web formalism, the potential is smoothed with a Gaussian of standard deviation $\sigma_\mathcal{N}$ Mpc/$h$ before the classification. The full segmentation procedure thus depends on three parameters that are: $\sigma_\mathcal{N}$, the Gaussian smoothing scale, $N_\mathrm{g}$, the total number of grid cells and $\lambda_\mathrm{th}$, the threshold for the eigenvalues of the tidal tensor. Both $N_\mathrm{g}$ and $\sigma_\mathcal{N}$ are related to a smoothing effect of the fields and can be combined in an effective smoothing scale defined as ${R_\mathrm{eff}}^2 = \left(L/N_\mathrm{g}\right)^2 + {\sigma_\mathcal{N}}^2$. The choice of $N_\mathrm{g}$ represents a trade-off between the minimum scales involved in the analysis and the resolution of the simulation containing, in the present case, $512^3$ particles. To be able to probe non-linear scales up to around $0.5$ $h$/Mpc, we find that $N_\mathrm{g} = 360$ is a good choice, leading to a half-Nyquist frequency of $k_\mathrm{Nyq}/2 = 0.57$ $h$/Mpc, with $k_\mathrm{Nyq} = \pi N_\mathrm{g} / L$. In practice, we set $k_\mathrm{max} = 0.5$ $h$/Mpc, allowing to take into account both large-scales and non-linear ones without introducing bias induced by aliasing effects occurring when $k > k_\mathrm{Nyq}/2$. On the other hand, $\sigma_\mathcal{N}$ blurs out structures at scales below this value. Physically, the size of the structures (radii of nodes, widths of filaments and walls) spread over a few Mpc/$h$ \citep{Cautun2014}. If we want to describe accurately the structures and, still aiming at a target scale of $12.5$ Mpc/$h$ in configuration space (corresponding to a Fourier mode of $0.5$ $h$/Mpc), $R_\mathrm{eff}$ should be limited so that some voxels can be used to probe it. Consequently, for $R_\mathrm{eff}$ to be below $4$ Mpc/$h$ and with $N_\mathrm{g} = 360$, $\sigma_\mathcal{N}$ should be below $3$ Mpc/$h$. In practice, we adopt $\sigma_\mathcal{N} = 2$ Mpc/$h$, leading to an effective smoothing scale of $R_\mathrm{eff}=3.4$ Mpc/$h$. After assessing several reasonable values of $\sigma_\mathcal{N} \in \left[1.5, 2.5\right]$ Mpc/$h$, we find that the volume fractions are changed by less than a percent while the mass fractions are modified by around 3\%. The chosen fiducial value of the smoothing scale is also coherent with previous usage of Gaussian smoothing for continuous fields before applying Hessian-based classification methods \citep[e.g.][]{Hahn2007, Nexus, Cautun2014, Martizzi2018}.
        
\begin{figure}
            \centering
            \includegraphics[width=1\linewidth]{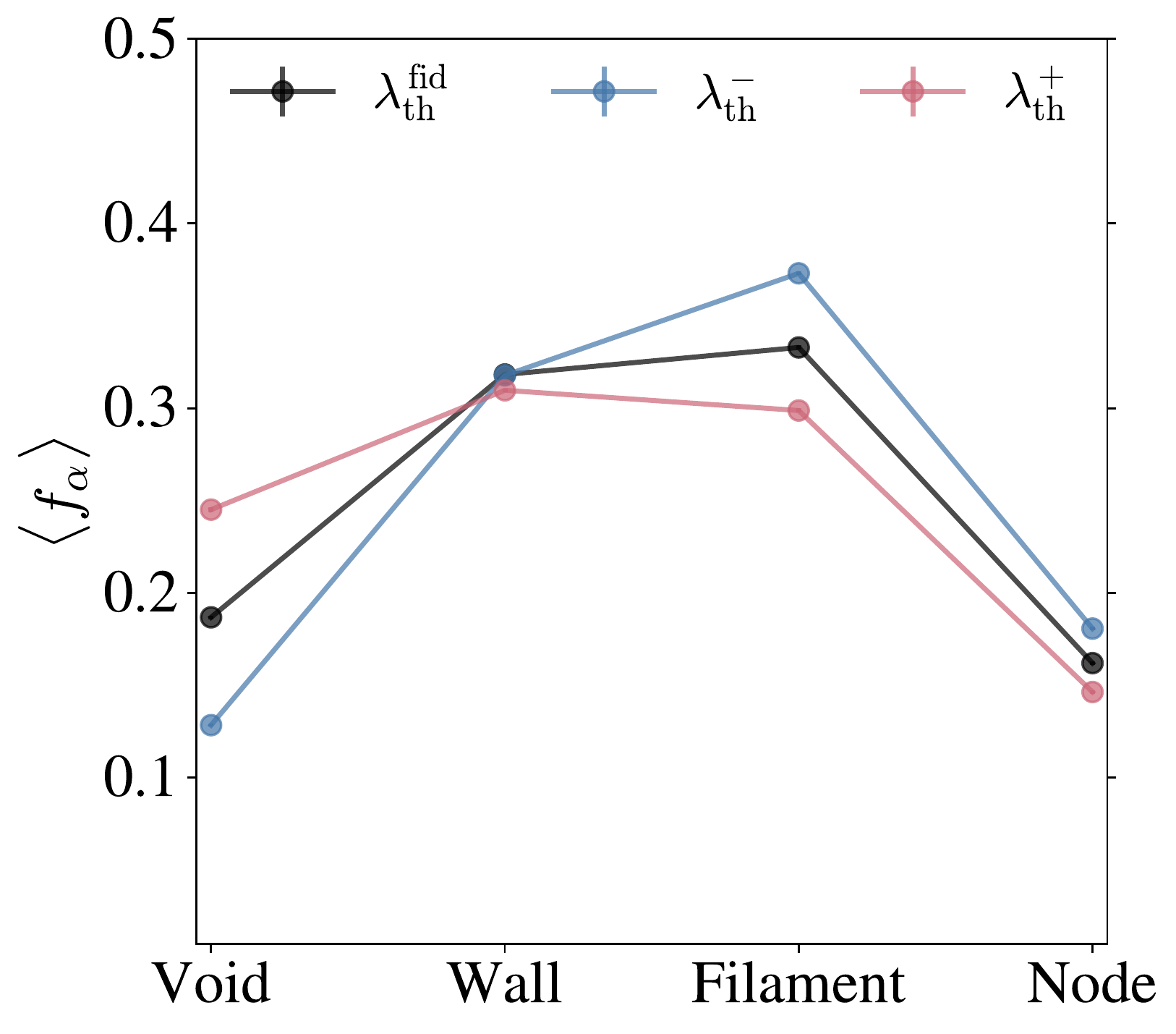}
            \caption{Averaged mass fractions $\langle f_\alpha \rangle$ in the different cosmic web environments for distinct values of $\{\lambda_\mathrm{th}^{-}, \lambda_\mathrm{th}^\mathrm{fid}, \lambda_\mathrm{th}^{+}\} = \{0.2, 0.3, 0.4\}$ of the T-web formalism.}
            \label{fig:fractions_lambda}
        \end{figure}

        The $\lambda_\mathrm{th}$ parameter, however, has a larger effect on the classification of environments, both in terms of volume and mass fractions \citep{Forero-Romero2009}. The impact of this parameter is illustrated in Fig.~\ref{fig:fractions_lambda} showing the averaged mass fractions $\langle f_\alpha \rangle$ for each cosmic web environment as drawn by the T-web formalism with three values of $\lambda_\mathrm{th}$ that are $\{\lambda_\mathrm{th}^{-}, \lambda_\mathrm{th}^\mathrm{fid}, \lambda_\mathrm{th}^{+}\} = \{0.2, 0.3, 0.4\}$. The fiducial value of $0.3$ corresponds roughly to the threshold at which voids percolate for the cosmological volume of $L^3 = 1$ (Gpc/$h$)$^3$ \citep{Forero-Romero2009}.
        Figure~\ref{fig:fractions_lambda} reports the mass fractions obtained with these three values of $\lambda_\mathrm{th}$ showing that an increasing number of particles is associated with voids and less with filaments and nodes when the threshold is increased. More quantitatively, varying $\lambda_\mathrm{th}$ from $0.2$ to $0.4$ yields a factor of two between the obtained mass fractions in voids.
        
        In order to derive constraints from environments that are robust to the identification of the environments, and hence indirectly to changes in the definitions of the various web-finder methods \citep[see e.g.][]{Libeskind2017}, we embed both $\sigma_\mathcal{N}$ and $\lambda_\mathrm{th}$ as nuisance parameters in the analysis such that all presented cosmological constraints are marginalised over it.
        
        \subsection{Cosmological sensitivity of the classification}
        
\begin{figure*}
            \centering
            \includegraphics[width=.49\linewidth]{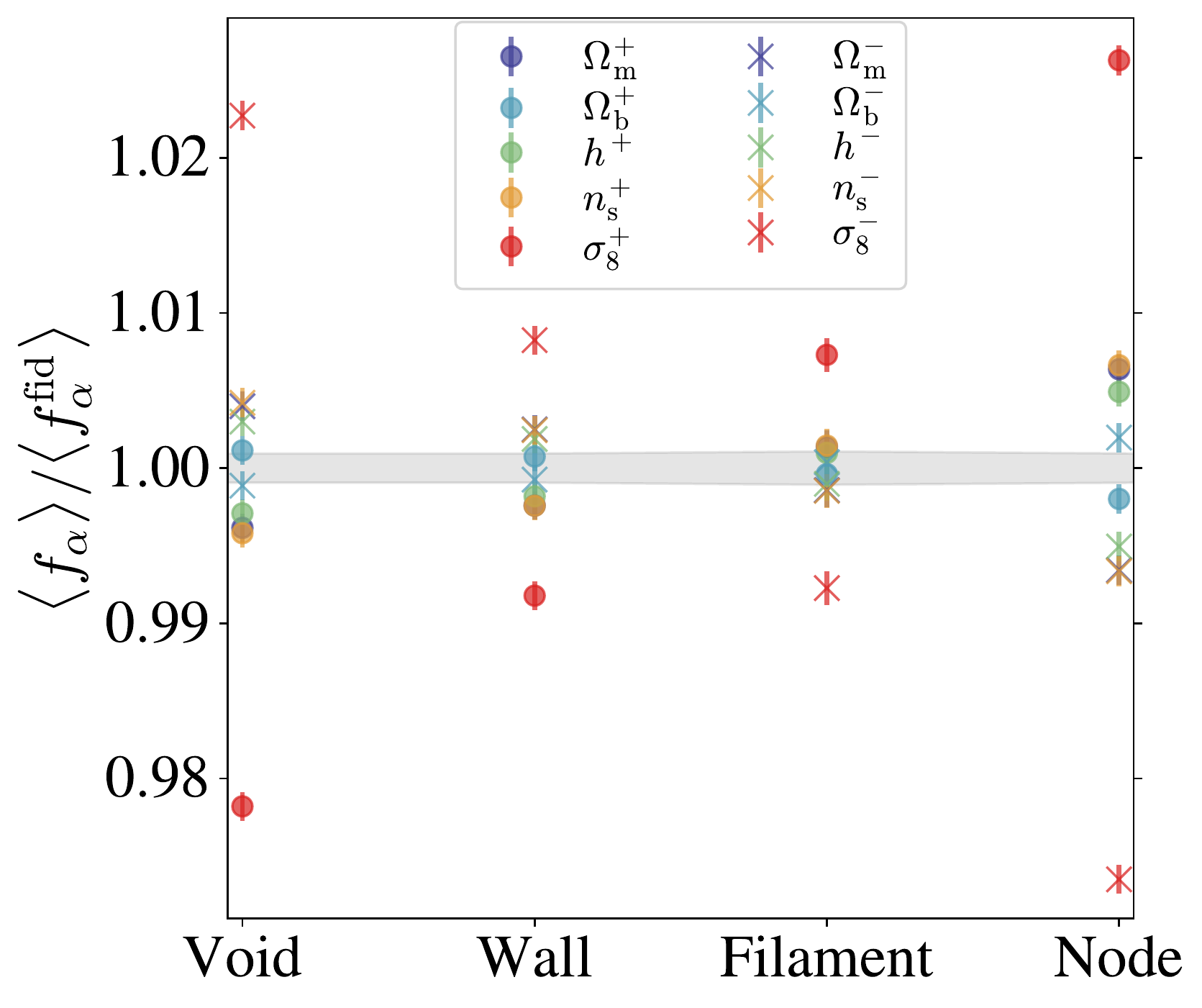}
            \includegraphics[width=.49\linewidth]{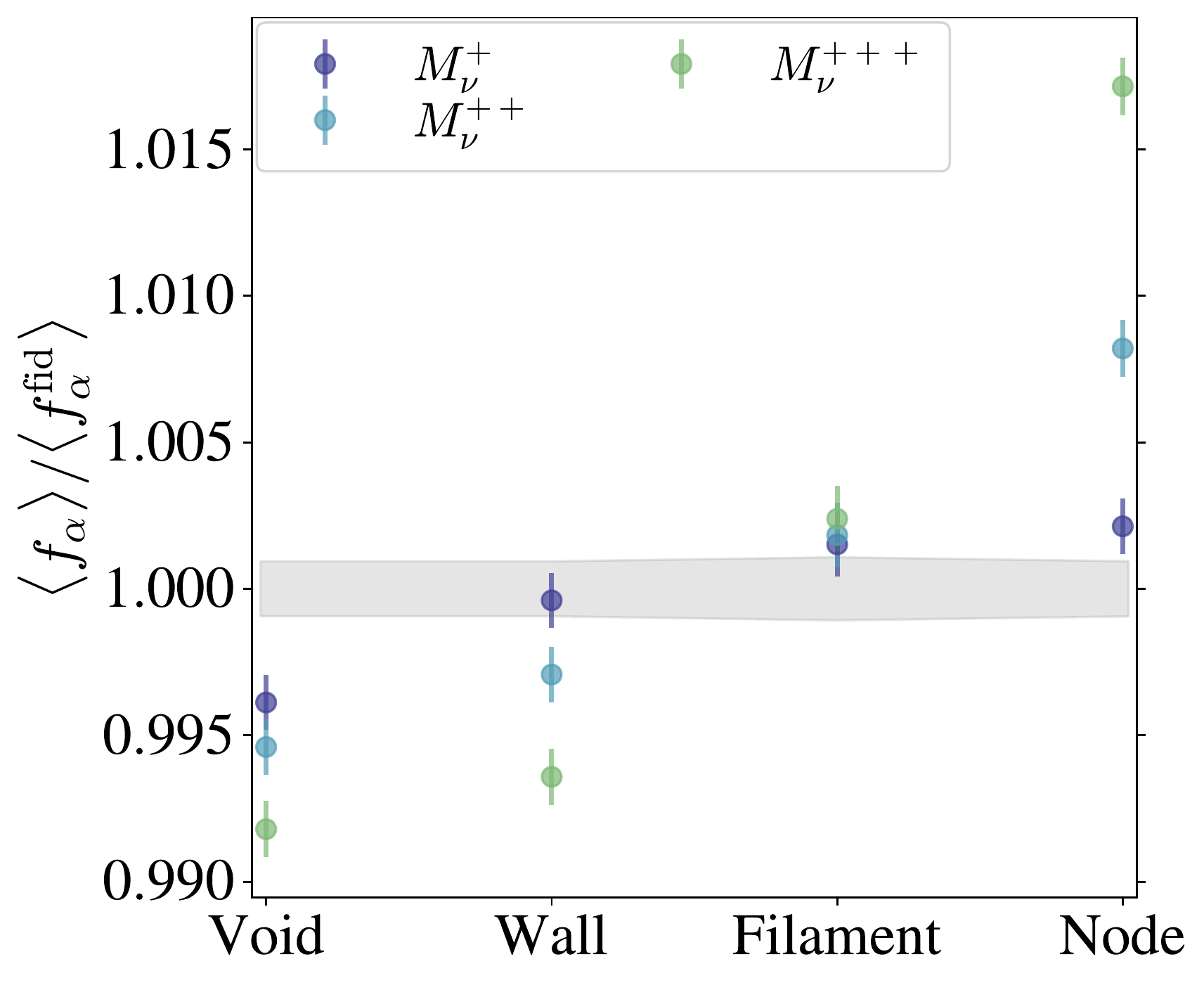}
            \caption{\textit{(left)} Ratio between mass fractions in the cosmic web environments when varying $\Omega_\mathrm{m}, \Omega_\mathrm{b}, h, n_\mathrm{s}, \sigma_8$ and mass fractions obtained with fiducial cosmology for an eigenvalue threshold of $\lambda_\mathrm{th}^\mathrm{fid}=0.3$. Points are centred on the average over the $N_\mathrm{deriv} = 500$ realisations for each cosmology and error bars show the $\pm 3\sigma$ interval. The grey shaded area corresponds to the $\pm 3\sigma$ interval of the fiducial cosmology fractions obtained from the $N_\mathrm{fid} = 7000$ realisations.
            \textit{(right)} Same when varying $M_\nu$. }
            \label{fig:fractions_real}
        \end{figure*}

        Figure~\ref{fig:fractions_real} shows the ratio between the averaged mass fractions in each environment when a cosmological parameter varies and the average obtained with the fiducial cosmology, $\langle f_\alpha \rangle/ \langle f^\mathrm{fid}_\alpha \rangle$. The error bars (represented as the bars around points and crosses, and as the grey shaded area for fiducial simulations) are the $3\sigma$ confidence intervals. Many cosmological parameters cause sizeable changes in these proportions and parameters related to matter density, like $\sigma_\mathrm{8}$ and $\Omega_\mathrm{m}$, are among those having the largest impact, together with $n_\mathrm{s}$. The most notable variations are induced by $\sigma_\mathrm{8}$, for which an increase (respectively decrease) is leading to a larger (respectively smaller) mass fraction in dense environments (nodes and filaments). This is in agreement with the definition of $\sigma_\mathrm{8}$ that is measuring how matter clusters at a scale of $8$ Mpc/$h$. The impact of neutrino mass, even though smaller in comparison, is significant with fraction ratios lying outside the $3\sigma$ confidence regions when varying this parameter. The right panel of Fig.~\ref{fig:fractions_real} shows that, similarly to the effect of $\sigma_\mathrm{8}$, increasing $M_\nu$ makes dense environments even denser.
        All these different effects observed in the mass fractions already suggest that each cosmological parameter has a different impact on the environments that will be even more refined when inspecting the information contained in the power spectra.
        
    \subsection{Power spectra in cosmic web environments}
    
        The auto power spectrum $P_{\alpha\alpha}(k)$ is defined as the covariance of Fourier modes of the overdensity field $\delta_\alpha$, with $\alpha \in \{\mathrm{v}, \mathrm{w}, \mathrm{f}, \mathrm{n}\}$ denoting voids, walls, filaments and nodes respectively. For an overdensity field $\delta_\alpha$, the power spectrum is given by
        \begin{equation}  \label{eq:PS_real}
            P_{\alpha\alpha}(k) \delta^{(3)}_\mathrm{D}(\bm{k}_1 + \bm{k}_2) = \frac{1}{\left(2\pi\right)^3} \langle \tilde{\delta}_\alpha(\bm{k}_1) \tilde{\delta}_\alpha(\bm{k}_2) \rangle,
        \end{equation} 
        with $k = \lVert \bm{k}_1 \rVert_2$, $\tilde{\delta}$ referring to the Fourier transform of $\delta$, and $\delta^{(3)}_\mathrm{D}$ is the Dirac delta distribution in $\mathbb{R}^3$.
        For simulations with $M_\nu > 0$, power spectra are computed based on the total matter density fields containing both dark matter and neutrino particles, $\delta_\mathrm{m} = f_\mathrm{CDM} \delta_\mathrm{CDM} + f_\nu \delta_\nu$, where $\delta_\nu$ is the neutrino field and $\delta_\mathrm{CDM}$ is the one from dark matter particles. The fraction $f_\nu$ is computed as $\Omega_\nu/\Omega_\mathrm{m}$ with $\Omega_\nu={M_\nu h^{-2}}/{93.14\, \mathrm{eV}}$ and $f_\nu + f_\mathrm{CDM} = 1$.
        
        The PCS smoothing scheme used to evaluate the overdensity fields $\delta_\alpha$ however deforms the shape of the estimated power spectra \citep{Jing2005}. We hence correct for this effect by first deconvolving the fields $\delta_\alpha$ through the application of the window function in Fourier space
        \begin{equation}
            W(k) = \left[ \prod_i \left(1 - \frac{4}{3} s_i + \frac{2}{5} s_i^2 - \frac{4}{315} s_i^3 \right) \right]^{-1},
        \end{equation}
        with $s_i = \sin\left(\pi k_i /2 k_\mathrm{Nyq}\right)$ and $k_\mathrm{Nyq}$ the Nyquist frequency.
        Finally, because of the discrete nature of the input, namely the dark matter particles (or the neutrinos in massive neutrino simulations), we also subtract the shot noise contribution from power spectra estimated with Eq.~\eqref{eq:PS_real}. Even though the number of particles is very large and we expect the shot noise contribution to be small at the scales of interest, auto-spectra $P_{\alpha\alpha}$ (including also $P_\mathrm{mm}$) are subtracted by the quantity $1/\bar{n}_\alpha$ where $\bar{n}_\alpha = N_\alpha / L^3$. In massive neutrino simulations, the shot noise is removed from the two weighed discrete contributions of the dark matter and the neutrino particles. This latter is however expected to be small on the overall power spectra, because of the $f_\nu^2$ weight which is small in the explored values of $M_\nu$.
        
\begin{figure}
            \centering
            \includegraphics[width=1\linewidth]{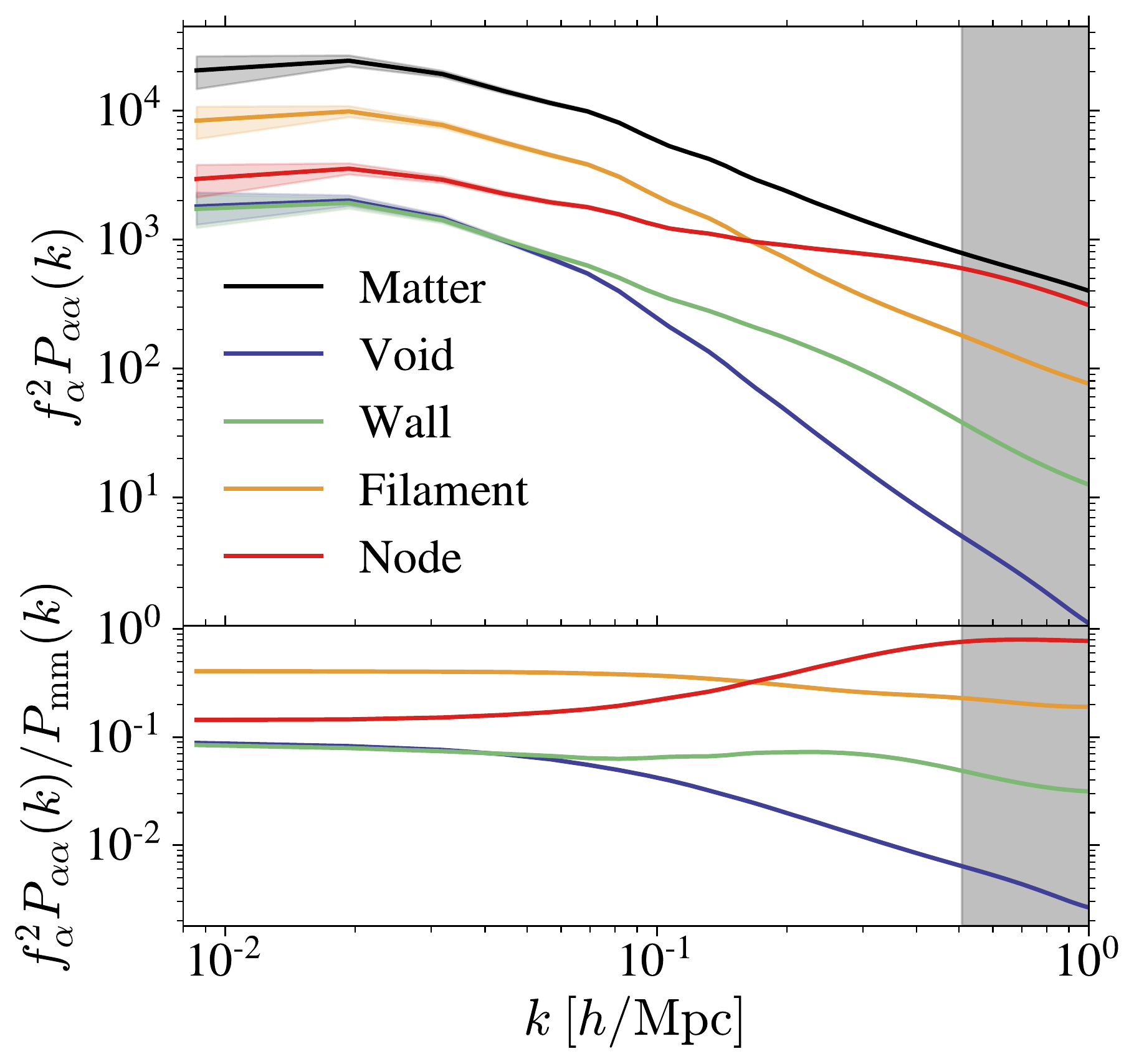} \\
            
            \caption{The top panel shows the real-space normalised power spectra $f_\alpha^2 P_{\alpha\alpha}(k)$ obtained for each environment $\alpha \in \{\mathrm{v}, \mathrm{w}, \mathrm{f}, \mathrm{n}\}$ compared to the one from all dark-matter particles in black. Solid lines are the average over the $N_\mathrm{fid}=7,000$ fiducial simulations and the shaded areas are the $1$-$\sigma$ intervals. The grey area depicts the range of $k>k_\mathrm{max}$ excluded from the analysis. In the bottom panel are found the ratios between the averaged normalised spectra in each environment and the matter-matter one (black from the top panel).}
            \label{fig:spectra_real}
        \end{figure}

        The auto-spectra are computed with a spectral bin size $\dd k = 2 k_\mathrm{f}$ with $k_\mathrm{f} = 2\pi N_\mathrm{g}/L \simeq 0.0126$ $h$/Mpc. For a maximum scale of $k_\mathrm{max} = 0.5$ $h$/Mpc, it yields $40$ bins in Fourier space for each environment, as well as for the matter spectra.
        The various auto-spectra $P_{\alpha\alpha}(k)$ are shown in Fig.~\ref{fig:spectra_real} in their normalised versions $f_\alpha^2 P_{\alpha\alpha}(k)$ to better see the contributions of each environment to the overall matter spectrum $P_\mathrm{mm}(k)$. 
        We qualitatively observe different shapes with $P_\mathrm{ff}(k)$ that globally looks like a shift of the matter one, emphasised in the bottom panel where are shown the ratios to $P_\mathrm{mm}(k)$. Node, void and wall auto-spectra however show different shape dependencies. In particular, we note that, at large-scales ($k<0.13$ $h$/Mpc), the dominant contribution is associated with filaments while most of the power is contained in nodes at smaller scales. At large scales also, $P_\mathrm{vv}(k)$ and $P_\mathrm{ww}(k)$ share similar amplitudes but the former quickly decreases when $k > 0.05$ $h$/Mpc with respect to all other environments. These different $k$-dependencies reflect the dissimilar statistical distributions of the matter in several cosmic web environments.
        
\section{Cosmological information content of cosmic web environments}  \label{sect:analysis}

    \subsection{Fisher formalism}
    
        Considering a set of model parameters $\bm{\theta} \in \mathbb{R}^d$ (in our case, cosmological parameters), we assume that the vector $\bm{s} \in \mathbb{R}^n$ is a statistic built from an observable (here, the binned power spectra drawn from the overdensity fields) following a Gaussian distribution $\bm{s} \sim \mathcal{N}( \bar{\bm{s}}, \bm{\Sigma})$. Its log-likelihood can hence be written
        \begin{equation} \label{eq:fisher_LL}
            \log p(\bm{s} \given \bm{\theta}) = -\frac{1}{2} (\bm{s} - \bar{\bm{s}})\tran \bm{\Sigma}^{-1} (\bm{s} - \bar{\bm{s}}) - \frac{1}{2} \log |\bm{\Sigma}| + \mathrm{const.},
        \end{equation}
        where the constant comes from the normalisation of the distribution.
        A common way to quantify the information carried by $\bm{s}$ on $\bm{\theta}$ is to use the Fisher information matrix $\bm{I}(\bm{\theta})$. From the Fréchet-Darmois-Cramér-Rao inequality, its inverse $\bm{I}(\bm{\theta})^{-1}$ corresponds to a lower-bound on the variance of any unbiased estimator drawn from $\bm{s}$ hence assessing the efficiency of the representation. Elements of the Fisher matrix are defined as the variance of the derivative of the log-likelihood, namely
        \begin{equation} \label{eq:FIM}
            \left[ \bm{I}(\bm{\theta}) \right]_{i,j} = \mathbb{E}_\theta \left[ \left(\frac{\partial \log p(\bm{s} \given \bm{\theta})}{\partial \theta_i}\right)\tran \left( \frac{\partial \log p(\bm{s} \given \bm{\theta})}{\partial \theta_j} \right) \right],
        \end{equation}
        which can also be written in terms of the second derivative of the log-likelihood under some smoothness constraints (which are fulfilled in the Gaussian case),
        \begin{equation}  \label{eq:FIM2}
            \left[ \bm{I}(\bm{\theta}) \right]_{i,j} = -\mathbb{E}_\theta \left[ \frac{\partial^2 \log p(\bm{s} \given \bm{\theta})}{\partial \theta_i\partial \theta_j} \right],
        \end{equation}
        where $\mathbb{E}_\theta$ is the expectation taken over the distribution $p(\bm{s} \given \bm{\theta})$.
        This latter equation intuitively explains how the amount of information is measured. A sharp log-likelihood around $\bm{\theta}$ implies a huge increase with small changes of the parameters, making the statistic $\bm{s}$ very sensitive to variations $\dd \theta$. On the other hand, a weakly-curved log-likelihood with a locally flat behaviour advocates for a poor description since its sensitivity with changes in the parameters is small.
        Under the Gaussian assumption described above and by further considering a covariance matrix $\bm{\Sigma}$ independent from cosmological parameters $\bm{\theta}$, mainly because this contribution is expected to be small and source of underestimation of errors \citep{Carron2013, Kodwani2019}, it yields
        \begin{equation} \label{eq:FIM_gaussian}
            \left[ \bm{I}(\bm{\theta}) \right]_{i,j} = \left(\frac{\partial \bar{\bm{s}}}{\partial \theta_i}\right)\tran \bm{\Sigma}^{-1} \left(\frac{\partial \bar{\bm{s}}}{\partial \theta_j}\right).
        \end{equation}
        The non-linear operation of the inversion to compute the precision matrix $\bm{\Sigma}^{-1}$ actually leads to a biased estimate, even though the covariance may be computed using the classical unbiased estimation. Still under the previously-established Gaussian assumption, the unbiased estimate of the precision matrix is given by \citep{Kaufman1964, Hartlap2007}
        \begin{equation} \label{eq:precision}
            \bm{\Sigma}^{-1} = \frac{N_\mathrm{fid} - n - 2}{N_\mathrm{fid} - 1} \, \hat{\bm{\Sigma}}^{-1},
        \end{equation}
        where $N_\mathrm{fid}$ is the number simulations at the fiducial cosmology, $n$ is the length of the summary statistics vector $\bm{s}$ and $\hat{\bm{\Sigma}}=\left(\bm{s} - \bar{\bm{s}} \right) \left(\bm{s} - \bar{\bm{s}} \right)\tran / \left(N_\mathrm{fid}-1\right)$ is the unbiased estimate of the covariance matrix.
        
        In our study, the partial derivatives of the summary statistics with respect to parameters of the model can be computed numerically using the variations of cosmologies provided by the Quijote suite of simulations. Considering the set of studied cosmological parameters $\{ \Omega_\mathrm{m}, \Omega_\mathrm{b}, h, n_\mathrm{s}, \sigma_\mathrm{8}\}$, we can estimate
        \begin{equation} \label{eq:derivatives}
        \frac{\partial \bar{\bm{s}}}{\partial \theta_i} \simeq \frac{\bar{\bm{s}}(\theta_i + \dd \theta_i) - \bar{\bm{s}}(\theta_i - \dd \theta_i)}{2 \dd \theta_i}. 
        \end{equation}
        In the case of massive neutrino simulations, $M_\nu > 0$ with a fiducial value at $0.0$ eV. For this parameter, we thus cannot rely on Eq.~\eqref{eq:derivatives} and instead estimate the derivative using the four-point forward approximation
        \begin{equation} \label{eq:derivatives_neutrinos}
        \frac{\partial \bar{\bm{s}}}{\partial M_\nu} \simeq \frac{ \bar{\bm{s}}(4 M_\nu^+) - 12 \bar{\bm{s}}(2 M_\nu^+) + 32 \bar{\bm{s}}(M_\nu^+) - 21\bar{\bm{s}}(M_\nu = 0.0)}{12 M_\nu^+}.
        \end{equation}
        Because massive neutrino simulations in Quijote are initialised using the Zel'dovich approximation and fiducial ones using the second order Lagrangian perturbation theory, the quantity $\bar{\bm{s}}(M_\nu = 0.0)$ is computed using the fiducial simulations initialised with the Zel'dovich approximation also.
        In all the presented results, if not mentioned otherwise, the numerical estimation of derivatives and covariances have been respectively made with $N_\mathrm{deriv}=500$ and $N_\mathrm{fid}=7000$ realisations. In Appendix \ref{appendix:convergence_fisher}, we discuss the impact of these numbers and assess the numerical stability of the results.

    \subsection{Information content of power spectra in cosmic web environments}
    
        \begin{figure}
            \centering
            \includegraphics[width=1\linewidth]{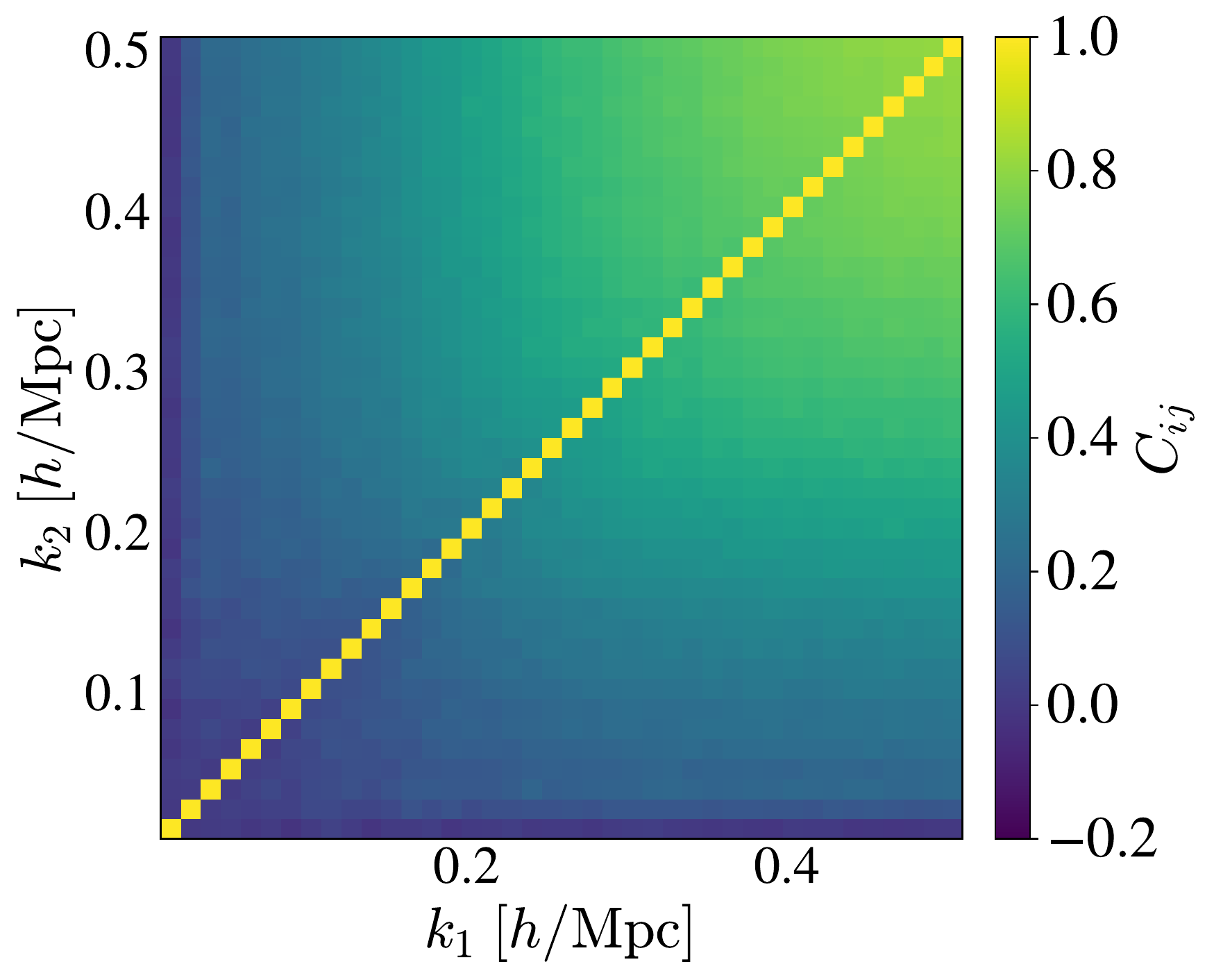}
            \includegraphics[width=1\linewidth]{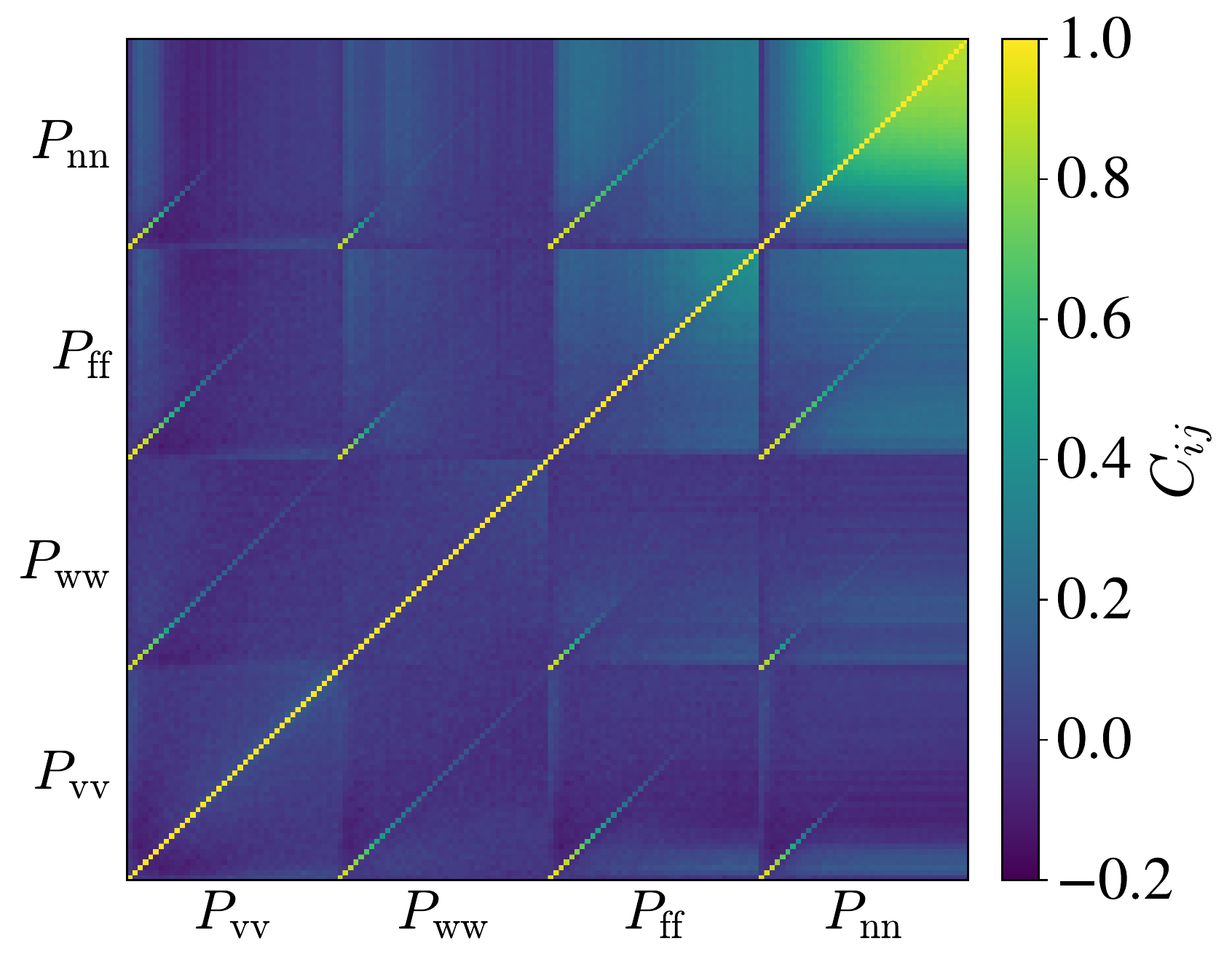}
            \caption{\textit{(top)} Correlation coefficients $C_{ij}$ for the matter power spectrum. \textit{(bottom)} Same for $P_{\alpha\alpha}(k)$ coefficients extracted from the several environments. Each sub-matrix goes from $k=0.1$ $h$/Mpc to $k=0.5$ $h$/Mpc.}
            \label{fig:correlation_real}
        \end{figure}
        
        \begin{figure*}
            \centering
            \includegraphics[width=1.0\linewidth]{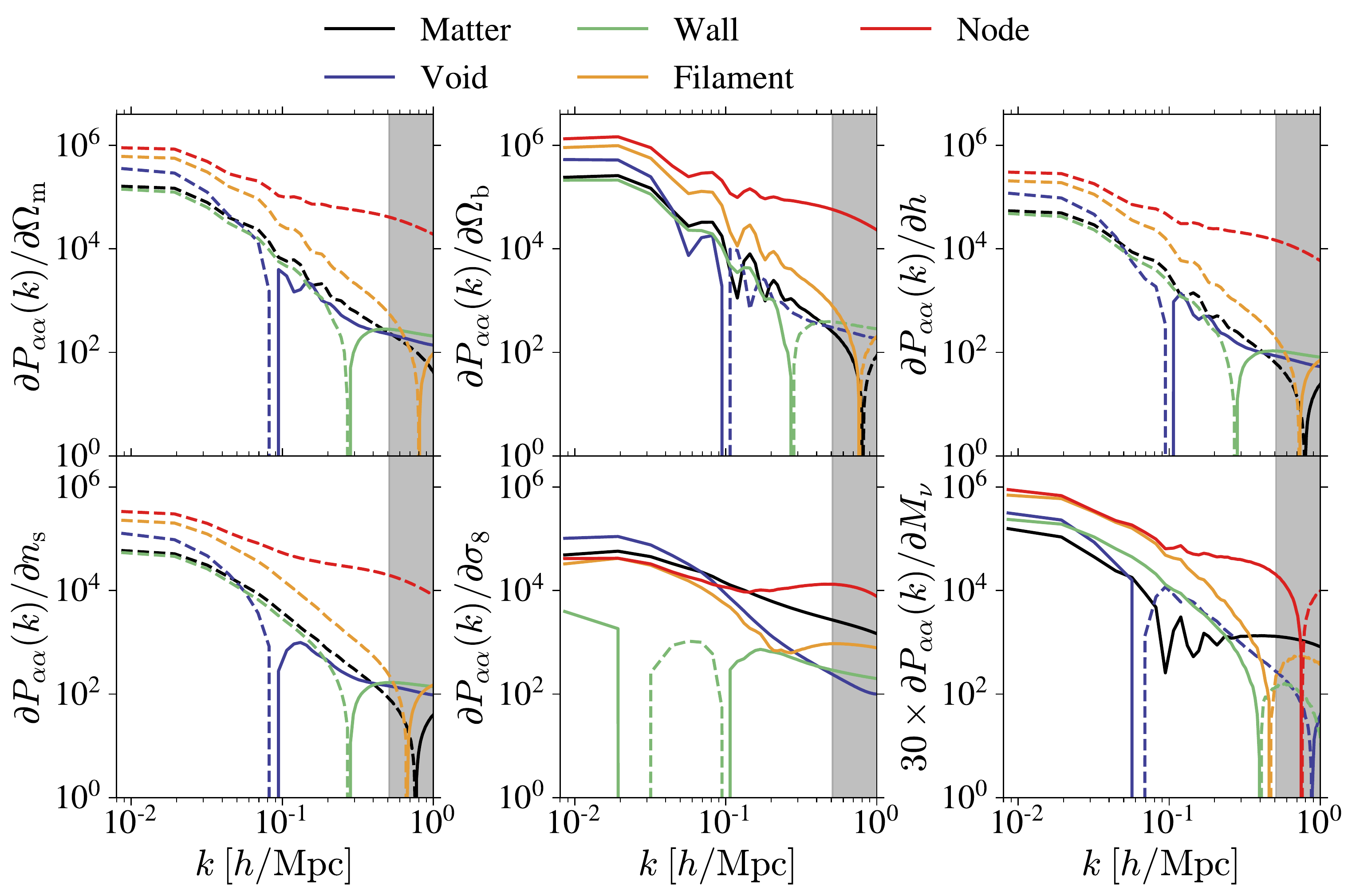}
            \caption{Partial derivatives $\partial P_{\alpha\alpha}(k) / \partial \theta_i$ for the different environmental auto-spectra and for each studied parameters $\{\Omega_\mathrm{m}, \Omega_\mathrm{b}, h, n_\mathrm{s}, \sigma_\mathrm{8}, M_\nu\}$. Dashed (respectively solid) lines correspond to negative (respectively positive) values of the derivative. The grey area depicts the range of $k>k_\mathrm{max}$ excluded from the analysis.}
            \label{fig:derivatives_real}
        \end{figure*}
        
        The two key ingredients of the Fisher-based quantification of information appear in Eq.~\eqref{eq:FIM_gaussian} as the covariance matrix $\bm{\Sigma}$ and the partial derivatives of the statistic with respect to the cosmological parameters.
        
        Figure~\ref{fig:correlation_real} plots a proxy of the first ingredient through the normalised version of the covariance matrix, namely the correlation matrix $\bm{C}$, whose elements are defined as
        \begin{equation}  \label{eq:corr_coefs}
            C_{ij} = \frac{\Sigma_{ij}}{\sqrt{\Sigma_{ii} \Sigma_{jj}}}.
        \end{equation}
        The first striking observation when inspecting the correlation matrix for $P_\mathrm{mm}$ in the top panel of Fig.~\ref{fig:correlation_real} is that it quickly becomes highly non-diagonal with correlation coefficients $C_{k_1, k_2} = 0.5$ at scales of $\sim 0.3$ $h$/Mpc. Such high couplings between scales are expected at low redshifts, Fourier modes being more and more correlated with time, as a result of the non-linear evolution of the matter distribution \citep[see][]{Blot2015}.
        These non-diagonal terms intrinsically reduce the representative power of the matter power spectrum to constrain the cosmological parameters, independently of how it varies with these latter. 
        The power spectra in different environments, in the bottom panel of Fig.~\ref{fig:correlation_real}, display less off-diagonal cross-correlation coefficients of high values, except in the node-node case. Indeed, Fourier modes of $P_\mathrm{nn}$ are highly correlated with values $C_{k_1 k_2} \sim 0.5$ for $k_1,k_2 \sim 0.2$ $h$/Mpc, which is a signature of the highly non-linear environment it represents. Other environments display lower values of the correlation coefficients at similar scales, with $\sim10^{-1}$ for filaments, $\sim10^{-2}$ for walls and $\sim5\times 10^{-2}$ for voids. Less and less correlations between modes are hence observed in environments with decreasing $\lvert \delta \rvert$ from non-linear regions like nodes and filaments, then mildly non-linear voids and finally walls having the closest-to-zero distribution of densities being consequently the environment exhibiting the least coupling between Fourier modes, as can be visually appreciated from Fig.~\ref{fig:correlation_real}.
        
        Figure~\ref{fig:derivatives_real} shows the second ingredient of the Fisher-forecast formalism: the partial derivatives for all the studied statistics. Compared to $P_\mathrm{mm}$, the spectra drawn from the cosmic web environments show different patterns in the derivatives, probing broader ranges of amplitudes and exhibiting different features. Taking the example of the $\Omega_\mathrm{m}$ parameter in the top left panel, the change of sign occurs at different scales for each environment, and seems to follow the order of average density, namely from voids to nodes. This pattern is also observed for other parameters like $\Omega_\mathrm{b}$, $h$, $n_\mathrm{s}$ or $M_\nu$. The similar $k$-dependencies observed between the partial derivatives of the matter power spectrum exhibits the limitation of this statistic to discriminate between effects of different cosmological parameters. Power spectra in the cosmic web environments, by showing various dependencies suggest that they bring different information on the set of cosmological parameters, and that, when combined all together, they may break degeneracies and allow to put tighter constraints on the underlying cosmological model. 
        
\begin{figure*}
            \centering
            \includegraphics[width=.85\linewidth]{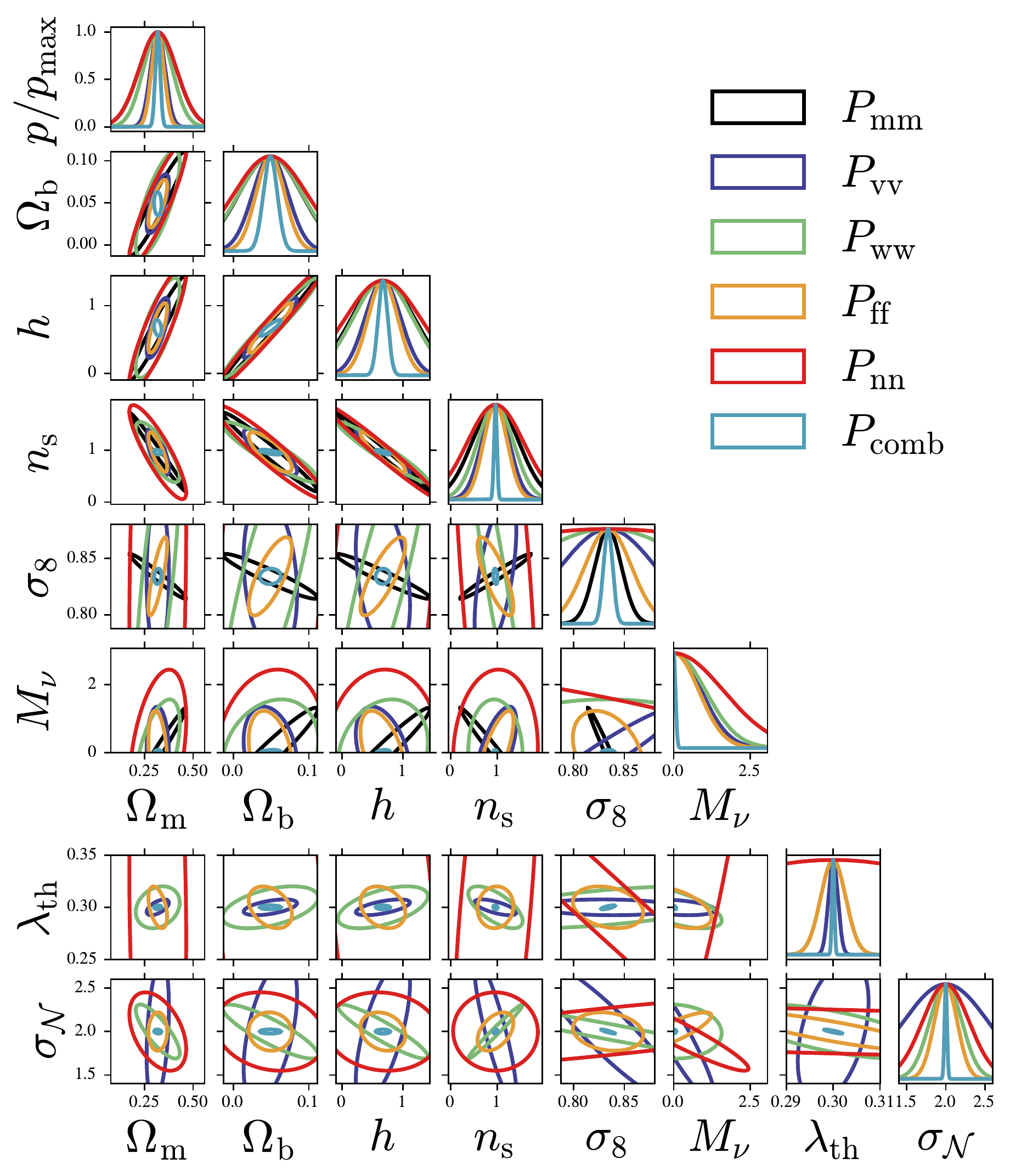}
            \caption{$1\sigma$ confidence ellipses for all the pairs of cosmological  $\left( \Omega_\mathrm{m}, \Omega_\mathrm{b}, h, n_\mathrm{s}, \sigma_\mathrm{8}, M_\nu \right)$ and nuisance ($\lambda_\mathrm{th}$) parameters obtained from the matter power spectrum, or the one from the different environments and their combination in real space. On the diagonal are shown the normalised probability density functions for each parameter.}
            \label{fig:corner_real}
        \end{figure*}

        One way to quantify the information gained by using the power spectra in the environments is to compute the marginalised $1\sigma$ confidence ellipses that can be obtained from the Fisher information matrices \eqref{eq:FIM_gaussian}. These latter are shown in Fig.~\ref{fig:corner_real} when using the matter power spectrum or the one from each environment either individually or combined all together. Table \ref{tab:constraints_real} gathers the marginalised $\sigma_{\theta_i}$ constraints obtained in the different cases defined as
        \begin{equation} \label{eq:margin_constraint}
            \sigma_{\theta_i} = \frac{1}{\sqrt{\left[\bm{I}(\bm{\theta})^{-1}\right]_{i,i}}},
        \end{equation}
        where $I_{i,i}$ is the $i\textsuperscript{th}$ diagonal element of the Fisher information matrix.
        The corner plot of Fig.~\ref{fig:corner_real} exhibits the degeneracies among parameters in almost all panels with black elongated ellipses translating the limitation of $P_\mathrm{mm}$ to distinguish between the effect of varying one or the other parameter, already hinted in the shapes of the derivatives. This is for instance observed in the $M_\nu$--$\sigma_\mathrm{8}$ panel \citep[also reported and studied in previous works like][]{Villaescusa-Navarro2013, Villaescusa-Navarro2014, Peloso2015}, or the $\Omega_\mathrm{m}$--$\sigma_8$ one.
        When inspecting the ellipses obtained from the spectra computed in the  individual environments, we clearly distinguish different orientations for several parameters. It is for instance especially striking in the $M_\nu$--$\sigma_8$ plane where $P_\mathrm{ff}$ and $P_\mathrm{vv}$ are showing quasi-orthogonal ellipses, but also in the projected space involving $\sigma_8$ and $h$, $n_\mathrm{s}$ or $\Omega_\mathrm{b}$.
        This illustrates the complementary information delivered by the two-point statistic in cosmic web environments, which, when combined all together, tightens up the constraints as quantitatively shown in Table \ref{tab:constraints_real} with improvement factors over the matter power spectrum ranging from $2.9$ to $15.7$ for the five cosmological parameters considered and $15.2$ for the sum of neutrino mass.
        The different structures observed in the derivatives of Fig.~\ref{fig:derivatives_real} and the individually lower couplings between $k$ modes in the covariance matrices from the bottom panel of Fig.~\ref{fig:correlation_real} also translate into better constraints for some environments and parameters, except nodes, and $\sigma_\mathrm{8}$. The high level of correlation brought by the $P_\mathrm{nn}$ coefficients and the small range of amplitudes probed by their derivatives indeed lead to large errors on cosmological parameters when exploiting this environment alone. Without similar high couplings, the other environments perform either equivalently or better than the matter power spectrum in most cases. It is worthy noting that these constraints do not take into account any additional prior information coming from other measurements that could improve them even further, such as those from the Cosmic Microwave Background experiments \citep[CMB,][]{Planck2018}.
        
        The results presented above are obtained by including $\lambda_\mathrm{th}$ and $\sigma_\mathcal{N}$ as nuisance parameters. When fixing those to their fiducial values, we obtain the constraints reported in Table~\ref{tab:constraints_real_fixed} that are quasi-identical for $P_\mathrm{comb}$ on cosmological parameters $\Omega_\mathrm{m}$, $\Omega_\mathrm{b}$, $h$ and $n_\mathrm{s}$. Adding the nuisance parameters and marginalising over them mainly impacts $\sigma_8$ and $M_\nu$ where the improvement factors are respectively lowered down from $7.2$ and $24.3$ to $2.9$ and $15.2$. This is mainly due to the additional degeneracies that the two additional parameters induce, with for instance $\sigma_8$ and $\sigma_\mathcal{N}$ that have similar effects on the power spectra, mainly a shift (also visible from the corresponding panel in Fig.~\ref{fig:corner_real}). The obtained results from $P_\mathrm{comb}$ are hence showing good robustness with  these nuisance parameters, knowing their impact on the segmentation of the environments and on the derived statistics. Quantitatively, voids, walls, filaments, nodes and their combination are respectively leading to marginalised error over the threshold of $\sigma_{\lambda_\mathrm{th}} = \{0.0050, 0.0133, 0.0133, 0.1785, 0.0013\}$ and over the smoothing parameter of $\sigma_{\sigma_\mathcal{N}}= \{0.5405, 0.2061, 0.1463, 0.2988, 0.0160\}$, showing that both are well-constrained by the combination of environment-dependent spectra. 
        Even though the choice of $\lambda_\mathrm{th}$ and $\sigma_\mathcal{N}$ may influence the identified cosmic structures, the analysis of the classification parameters teaches us that it affects only partially the derived cosmological constraints. This is especially encouraging in the sense that it leaves room to other definitions of cosmic environments to be applied. Even though leading to different detected structures \citep[see e.g.][]{Libeskind2017}, we should end up with similar results at the constraints level.
        Interestingly, we also report that, either fixing the T-web formalism parameters or leaving them free, the filament environment provides the best individual constraints. When comparing Tables \ref{tab:constraints_real} and \ref{tab:constraints_real_fixed}, we see that filaments are indeed performing individually better than the matter power spectrum for most parameters, closely followed by the $P_\mathrm{vv}$ statistic. We notice that some environments are constraining particularly well some cosmological parameters, such as $M_\nu$ for voids, as theoretically expected and stated in previous works \citep[e.g.][]{Pisani2015, Massara2015, Kreisch2019} but also filaments which provides tight constraints on $\Omega_\mathrm{m}$ or $M_\nu$ compared to $P_\mathrm{mm}$.
        
        \begin{table*}
        \centering
        \caption{Marginalised 1-$\sigma$ constraints obtained from the power spectrum in different environments for all cosmological parameters in real space and considering $\lambda_\mathrm{th}$ and $\sigma_\mathcal{N}$ as nuisance parameters. $\sigma_{M_\nu}$ is in unit of eV.}
        \label{tab:constraints_real}
        
        \renewcommand{\arraystretch}{1.5}
        \smallskip
        \begin{adjustbox}{max width=1.0\textwidth,center}
        \begin{tabular}{c|cccccc}
            \toprule
            Statistics & $\sigma_{\Omega_\mathrm{m}}$ & $\sigma_{\Omega_\mathrm{b}}$ & $\sigma_{h}$ & $\sigma_{n_\mathrm{s}}$ & $\sigma_{\sigma_\mathrm{8}}$ & $\sigma_{M_\nu}$ \\
            \midrule
            $P_\mathrm{mm}$ & $0.0969$ & $0.0413$ & $0.5145$ & $0.5019$ & $0.0132$ & $0.8749$ \\
            
            $\color{void} P_\mathrm{vv}$ & $0.0381 \, (2.5)$ & $0.0234 \, (1.8)$ & $0.295 \, (1.7)$ & $0.2962 \, (1.7)$ & $0.0466 \, (0.3)$ & $0.8982 \, (1.0)$ \\
            $\color{wall} P_\mathrm{ww}$ & $0.0752 \, (1.3)$ & $0.0419 \, (1.0)$ & $0.4902 \, (1.0)$ & $0.3852 \, (1.3)$ & $0.0903 \, (0.1)$ & $1.0374 \, (0.8)$ \\
            $\color{filament} P_\mathrm{ff}$ & $0.0320 \, (3.0)$ & $0.0189 \, (2.2)$ & $0.2444 \, (2.1)$ & $0.2546 \, (2.0)$ & $0.0230 \, (0.6)$ & $0.8151 \, (1.1)$ \\
            $\color{node} P_\mathrm{nn}$ & $0.0971 \, (1.0) $ & $0.0481 \, (0.9) $ & $0.6142 \, (0.8)$ & $0.6004 \, (0.8)$ & $0.2006 \, (0.1)$ & $1.6178 \, (0.5)$ \\
            $\color{combination} P_\mathrm{comb}$ & $0.0126 \, (\bm{7.7})$ & $0.0093 \, (\bm{4.5})$ & $0.0793 \, (\bm{6.5})$ & $0.0319 \, (\bm{15.7})$ & $0.0046 \, (\bm{2.9})$ & $0.0575 \, (\bm{15.2})$ \\
            \bottomrule
        \end{tabular}
        \end{adjustbox}
        \end{table*}
        
        \begin{figure*}
            \centering
            \includegraphics[width=1.0\linewidth]{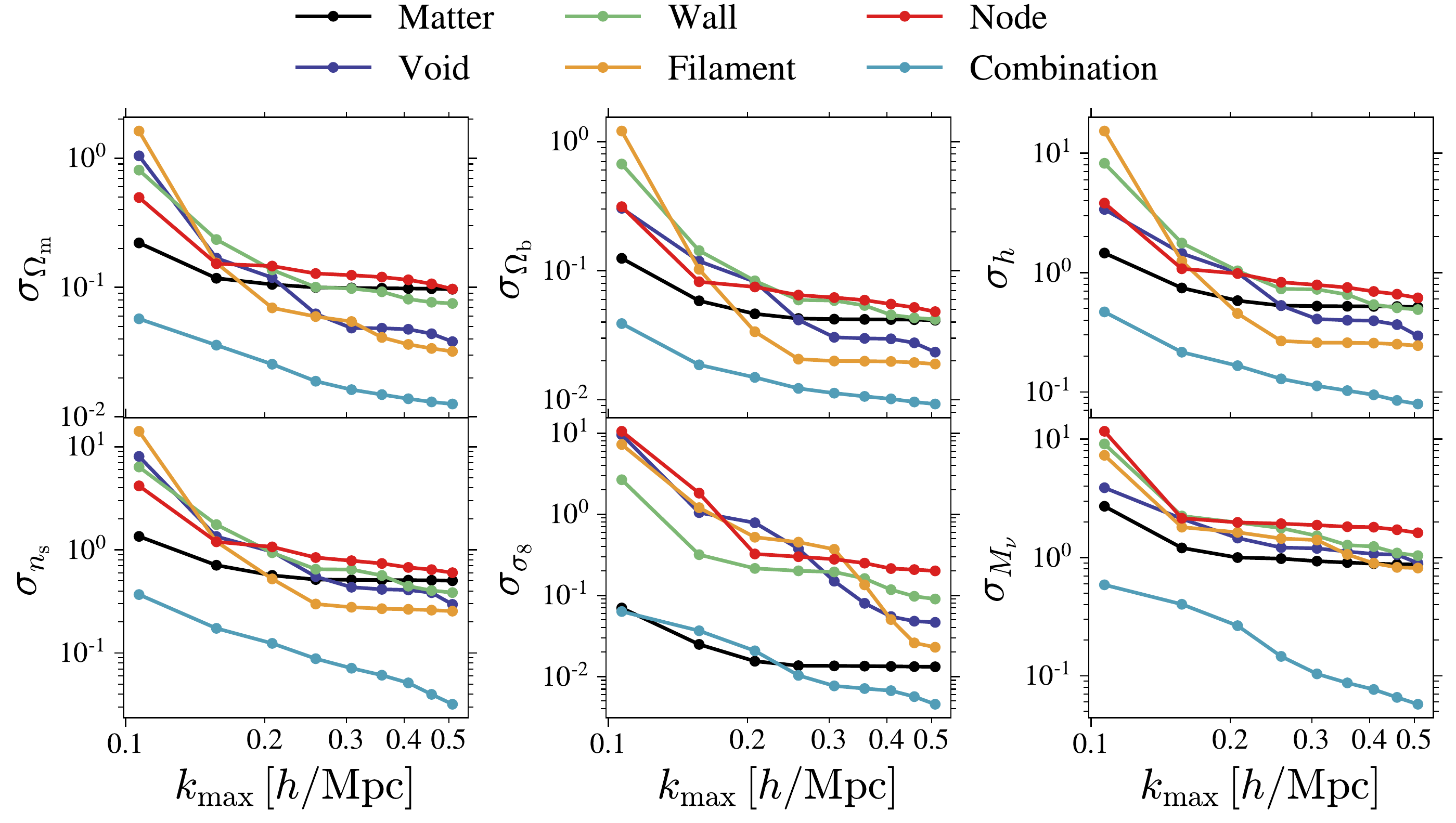}
            \caption{Evolution of the marginalised constraint $\sigma_{\theta_i}$ on cosmological parameters $\{\Omega_\mathrm{m}, \Omega_\mathrm{b}, h, n_\mathrm{s}, \sigma_8\}$ and the sum of neutrino mass $M_\nu$ with the maximum scale used for the Fisher analysis, $k_\mathrm{max}$. $\sigma_{M_\nu}$ is in unit of eV.}
            \label{fig:constraints_kmax_real}
        \end{figure*}
        
        \begin{figure}
            \centering
            \includegraphics[width=1\linewidth]{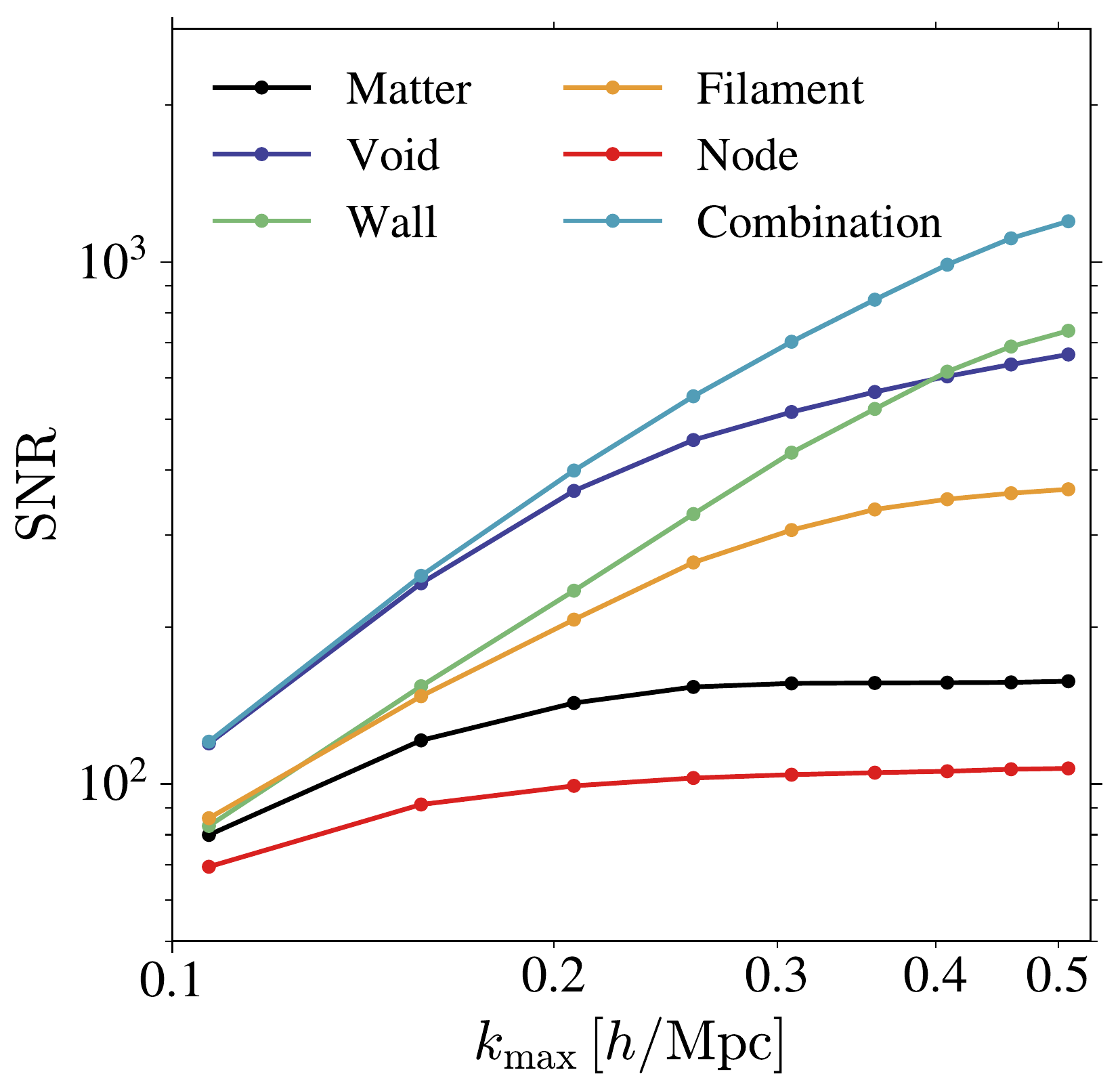}
            \caption{Evolution of the $\mathrm{SNR}$ with $k_\mathrm{max} \in \left[0.11, 0.5\right]$ Mpc/$h$ for the several studied statistics: the matter power spectrum, the power spectrum computed in the cosmic web environments, and the combination of environment-dependent spectra.}
            \label{fig:SNR_kmax_real}
        \end{figure}
        
        The marginalised constraints reported so far are obtained by including all the modes of power spectra spectra below $k_\mathrm{max}=0.5$ $h$/Mpc.
        Figure~\ref{fig:constraints_kmax_real} illustrates the evolution of $\sigma_{\theta_i}$ for each parameter and studied statistic with the value of the maximum scale involved to derive them $k_\mathrm{max} \in \left[0.1, 0.5\right]$ $h$/Mpc. The first conclusion we can draw is that the information extracted from $P_\mathrm{mm}$ saturates when $k_\mathrm{max}$ increases. This saturation, also pointed out by previous analyses \citep[e.g.][]{Takahashi2010, Blot2015, Chan2017} mostly comes from the degeneracies among parameters for this precise summary statistics which does not lead to any further improvement on the constraints at mildly non-linear scales when $k_\mathrm{max}>0.25$ $h$/Mpc. The smaller errors obtained in the filament and void environments are not observed at all scales. In particular, when restricting the analysis to $k_\mathrm{max} < 0.2$ $h$/Mpc, environments are not individually constraining better the set of cosmological parameters, even though their combination still leads to an improvement over the matter power spectrum analysis, for all considered values of $k_\mathrm{max}$.
        \par\bigskip
        An alternative quantity measuring the information carried by a statistic is given by the signal-to-noise ratio (hereafter $\mathrm{SNR}$) that describes the reachable accuracy of the statistic measurement given the covariance matrix. In general, the $\mathrm{SNR}$ of a summary statistic $\bm{s} \in \mathbb{R}^{n}$ is defined as
        \begin{equation}  \label{eq:SNR}
            \mathrm{SNR}(\bm{s}) = \sqrt{\bm{s}\tran \bm{\Sigma}^{-1} \bm{s}}\,,
        \end{equation}
        with $\bm{\Sigma}^{-1}$ the corresponding precision matrix defined by Eq.~\eqref{eq:precision}. Figure~\ref{fig:SNR_kmax_real} shows the evolution of the $\mathrm{SNR}$ for the matter and environment-dependent power spectra as a function of the maximum scale $k_\mathrm{max}$. We observe again the flattening when $k_\mathrm{max}$ approaches non-linear scales at $0.25$ $h$/Mpc for $\mathrm{SNR}(P_\mathrm{mm})$ \citep[also reported in][]{Angulo2008, Takahashi2010, Blot2015}. This feature is also observed for some spectra in environments like filaments but at higher $k$ values and all environment-dependent statistic are reaching a higher value of the $\mathrm{SNR}$, except for nodes. This latter saturates at an $\mathrm{SNR}$ $1.5$ times smaller than the matter power spectrum at $0.5$ $h$/Mpc, making it the lowest value among all environments, coherently with the findings of the Fisher forecast. Void is the environment performing best individually over a wide range of scales although overtaken by walls when $k_\mathrm{max} > 0.35$ $h$/Mpc.
        The shapes of the $\mathrm{SNR}$ evolution with $k_\mathrm{max}$ from the combination of environments also suggest that there is room left for a further increase when going to even smaller scales, $k_\mathrm{max} > 0.5$ $h$/Mpc, thanks to the breaking of degeneracies where the matter analysis will not be able to improve anymore. Quantitatively, the $\mathrm{SNR}$ obtained from the combination of environments is eight times higher than the one from $P_\mathrm{mm}$ at $k_\mathrm{max} = 0.5$ $h$/Mpc.
    
\section{Discussion} \label{sect:discussion}
    
    The obtained results show that using summary statistics derived after segmenting the matter distribution into the different components of the cosmic web (voids, walls, filaments and nodes) provides a better leverage on cosmological parameters in real space than the matter power spectrum analysis. The derived statistics in individual environments also exhibit less correlations between Fourier modes, except for the highly non-linear environments that are nodes. Recent analyses with alternative approaches like the marked power spectrum and the wavelet scattering transform (hereafter WST) were employed on the very same sets of simulations and of cosmological parameters in \cite{Massara2021} and \cite{Valogiannis2021} respectively. Alternatively, \cite{Bayer2021} propose an analysis using the information from the halo mass function (HMF) and the void size function (VSF) in real space. All three analyses, together with ours, consider a maximum scale of $k_\mathrm{max} = 0.5$ $h$/Mpc and can hence be naturally compared. The constraints obtained from these several statistics are summarised in Table \ref{tab:constraints_discussion}. The information contained in the power spectra bins in cosmic web environments provides the tightest constraints on $\Omega_\mathrm{m}$, $\Omega_\mathrm{b}$, $h$, and $n_\mathrm{s}$, and competitive values for constraints on $M_\nu$ and $\sigma_8$ which are however both best constrained by the coefficients derived from the WST. Compared to the partial information of extreme environments from voids and nodes performed in \cite{Bayer2021}, our environment-dependent statistics are able to tighten the constraints with improvement factors of $\{0.5, 4.1, 2.9, 3.1, 1.4, 1.7\}$ on the parameters $\{ \Omega_\mathrm{m}, \Omega_\mathrm{b}, h, n_\mathrm{s}, \sigma_\mathrm{8}, M_\nu \}$, hence performing better than the VSF+HMF statistic in all cases, except for $\Omega_\mathrm{m}$.
    
    These comparisons should however be nuanced in multiple aspects. First, all the forecasts are performed in real space, and some statistics could be deeply affected by the redshift-space distortions breaking the density field isotropy. This is for instance the case of the WST that requires adaption of the wavelets to be applied in this setting \citep{Valogiannis2021}, but also in our case, where the detection of the environment can be affected by the distortions, reducing the representative power of the environment-dependent statistics. Second, it is important to remind that the analysis we propose is the only one including extra nuisance parameters related to the cosmic web classification, $\lambda_\mathrm{th}$ and $\sigma_\mathcal{N}$, consequently increasing artificially the dimension of the parameter space and hence decreasing the constraints on the target parameters through additional degeneracies. As a fairer comparison, the last column of Table \ref{tab:constraints_discussion} shows the constraints we obtain when fixing the parameters of the classification method at their fiducial values.
    Finally, not only the obtained values of the constraints are important, but also the physical interpretability of the coefficients encoding the field information. 
    In that regard, the physical interpretability of the power spectra in cosmic web environments, being a simple two-point function, or that of the halo mass function and void size function are straightforward.
    
    \begin{table*}
    \centering
    \caption{Comparison of the marginalised 1-$\sigma$ constraints obtained from different Fisher forecasts using different statistics. $M$ stands for the marked power spectrum, $\mathrm{WST}$ for the wavelet scattering transform, HMF + VSF for halo mass function and void size function. In the $P_\mathrm{comb}$ column, ``Free'' refers to the constraints obtained when marginalising over the two nuisance parameters related to the segmentation scheme, $\lambda_\mathrm{th}$ and $\sigma_\mathcal{N}$, while ``Fixed'' are the ones when fixing those parameters at their fiducial values. $\sigma_{M_\nu}$ is in unit of eV.}
    \label{tab:constraints_discussion}
    
    \renewcommand{\arraystretch}{1.5}
    \smallskip
    \begin{adjustbox}{max width=1.0\textwidth,center}
    \begin{tabular}{c|ccccc}
        \toprule
        \multirow{2}{*}{Statistic} & \multirow{2}{*}{$M$} & \multirow{2}{*}{$\mathrm{WST}$} & \multirow{2}{*}{HMF + VSF} & \multicolumn{2}{c}{$P_\mathrm{comb}$} \\
        & & & & Free & Fixed \\
        Publication & \citeauthor{Massara2021} & \citeauthor{Valogiannis2021} & \citeauthor{Bayer2021} & \multicolumn{2}{c}{Bonnaire et al.} \\
        \midrule
        $\sigma_{\Omega_\mathrm{m}}$ & $0.013$ & $0.014$ & $0.006$ & $0.012$ & $0.012$ \\
        $\sigma_{\Omega_\mathrm{b}}$ & $0.010$ & $0.012$ & $0.037$ & $0.009$ & $0.009$ \\
        $\sigma_{h}$ & $0.098$ & $0.104$ & $0.23$ & $0.079$ & $0.078$ \\
        $\sigma{n_\mathrm{s}}$ & $0.048$ & $0.031$ & $0.100$ & $0.032$ & $0.031$ \\
        $\sigma_{\sigma_8}$ & $0.002$ & $0.001$ & $0.007$ & $0.005$ & $0.002$ \\
        $\sigma_{M_\nu}$ & $0.017$ & $0.008$ & $0.096$ & $0.058$ & $0.036$ \\
        \bottomrule
    \end{tabular}
    \end{adjustbox}
    \end{table*}

\section{Summary and conclusions} \label{sect:conclusions}

    In this work, we have carried out the first quantitative analysis of the cosmological information content of power spectra from the several cosmic web environments (nodes, filaments, walls, and voids). The derived statistics were computed from density fields associated with the environments identified through the eigenvalues of the Hessian matrix of the gravitational potential. Using the large suite of Quijote simulations, we performed a Fisher forecast by estimating numerically the partial derivatives and the covariance matrices of the extracted statistics in the non-linear regime with $k_\mathrm{max} = 0.5$ $h$/Mpc. We then compared the constraints on the cosmological parameters $\{\Omega_\mathrm{m}, \Omega_\mathrm{b}, h, n_\mathrm{s}, \sigma_\mathrm{8}, M_\nu\}$ derived from the cosmic web environments to those from the analysis of the matter power spectrum, marginalising over the two parameters of the web-finder method.
    From this analysis in real space, we report that:
    \begin{itemize}
        \item Environment-dependent spectra show different shape dependencies when varying cosmological parameters such as $M_\nu$, $\Omega_\mathrm{m}$ and $\sigma_\mathrm{8}$ with respect to each other and to the matter power spectrum. These variations originate from the intrinsic differences in densities and hence evolution histories of each environment, where the observed structures at $z=0$ are imprinted differently depending on the cosmology.
        \item Power spectra in void, wall and filament environments are less subject to mode coupling than the one computed from the matter for which overdense regions like nodes induce correlations between Fourier modes at small scales. As a result, voids and filaments for instance perform individually better than a matter power spectrum analysis for all cosmological parameters, except $\sigma_8$.
        \item The combination of power spectra in the environments leads to the breaking of some key degeneracies between parameters of the cosmological model which consequently tightens the constraints with improvement factors of $\{7.7, 4.5, 6.5, 15.7, 2.9, 15.2\}$ respectively on parameters $\{\Omega_\mathrm{m}, \Omega_\mathrm{b}, h, n_\mathrm{s}, \sigma_\mathrm{8}, M_\nu\}$ over the matter power spectrum. It also yields an $8$ times higher signal-to-noise ratio.
        \item The constraints obtained from the combination of environments are superior to the one of the matter power spectrum for the whole range of maximum scales analysed in the range $k_\mathrm{max} \in \left[0.1, 0.5\right]$ $h$/Mpc. 
        \item For the maximum scale involved in the analysis, $k_\mathrm{max}=0.5$ $h$/Mpc, the combination of environment-dependent spectra led to competitive, if not better in some cases, constraints compared to other state-of-the-art numerical analysis relying on the same set of simulations, with the advantage of being easily interpreted.
        \item The reported constraints are robust to variations of the two main parameters used to classify the environments, $\lambda_\mathrm{th}$ and $\sigma_\mathcal{N}$. The results are the same at the percent level for $\Omega_\mathrm{m}$, $\Omega_\mathrm{b}$ and $h$ while $n_\mathrm{s}$, $\sigma_8$ and $M_\nu$ respectively goes from $\{16.4, 7.2, 24.3\}$ to $\{15.7, 2.9, 15.2\}$ improvement factors over $P_\mathrm{mm}$.
    \end{itemize}
    
    \medskip
    
    In conclusion, we have shown that there is significantly more information contained in the density field when analysing the cosmic environments individually and using the combination of two-point statistics rather than when directly relying on the matter density field summarised by solely its power spectrum.
    The sizeable improvements in the constraints on all cosmological parameters brought by our environment-dependent analysis, even in the ideal case addressed in the present study, opens up the possibility to take advantage of the power spectra in environments for the optimal exploitation of future large galaxy redshift surveys such as the Dark Energy Spectroscopic Instrument \citep[DESI,][]{Levi2013} or Euclid \citep{Leureijs2011}. The present study however focused on the real-space constraints obtained using the information of cosmic web environments, alone and combined. Observations are carried in redshift-space in which the peculiar velocities of matter tracers distort the spatial distribution, especially in dense regions. Assessing the information gain in the redshift-space, where the multipole decomposition of the power spectrum already allows to tighten the constraints for the matter component, is needed and will be the purpose of a forthcoming paper (Bonnaire et al., in prep).
    Moreover, actual observations will provide us with biased tracers of the matter distribution, like halos or galaxies, instead of dark matter particles. Using statistics from cosmic web-dependent environments will necessitate to handle this complex relation between matter distribution and its tracers, through notably the inclusion of additional nuisance parameters related to biases effects.


\begin{acknowledgements}
The authors thank all members of the ByoPiC team\footnote{\url{https://byopic.eu/team}} for useful comments and discussions. We also thank the Quijote team for making their data publicly available and particularly F. Villaescusa-Navarro for his help and availability regarding the Quijote suite.
This research was supported by funding for the ByoPiC project from the European Research Council (ERC) under the European Union’s Horizon 2020 research and innovation program grant number ERC-2015-AdG 695561.
A.D. was supported by the Comunidad de Madrid and the Complutense University of Madrid (Spain) through the Atracción de Talento program (Ref. 2019-T1/TIC-13298).
T.B. acknowledges funding from the French government under management of Agence Nationalede la Recherche as part of the ``Investissements d’avenir'' program, reference ANR-19-P3IA-0001 (PRAIRIE 3IA Institute).
\end{acknowledgements}

\medskip

\bibliographystyle{aa}
\bibliography{thesis}

\begin{appendix}

\section{Stability and convergence analysis} \label{appendix:convergence_fisher}

\begin{figure}
        \centering
        \includegraphics[width=1\linewidth]{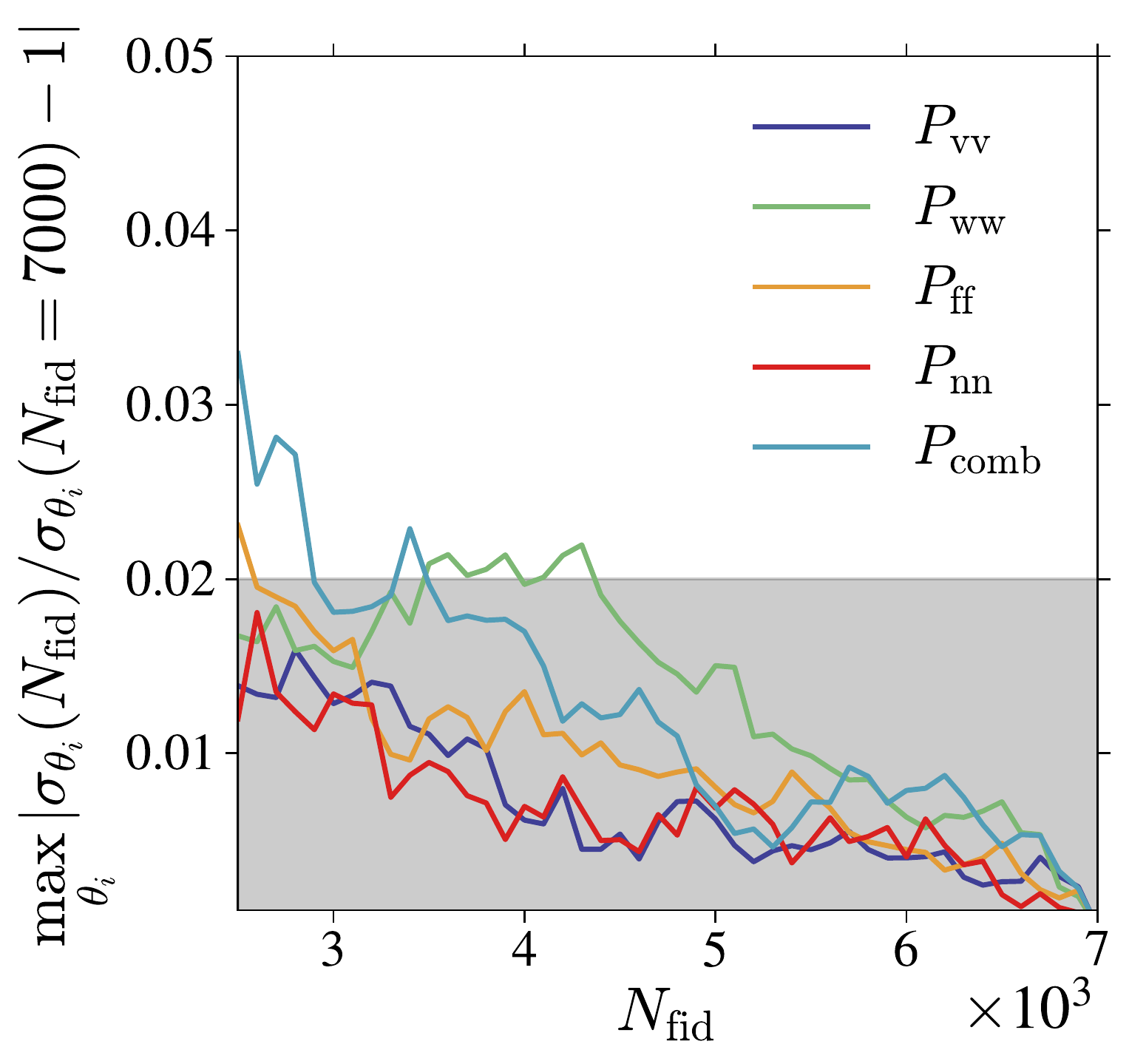}
        \includegraphics[width=1\linewidth]{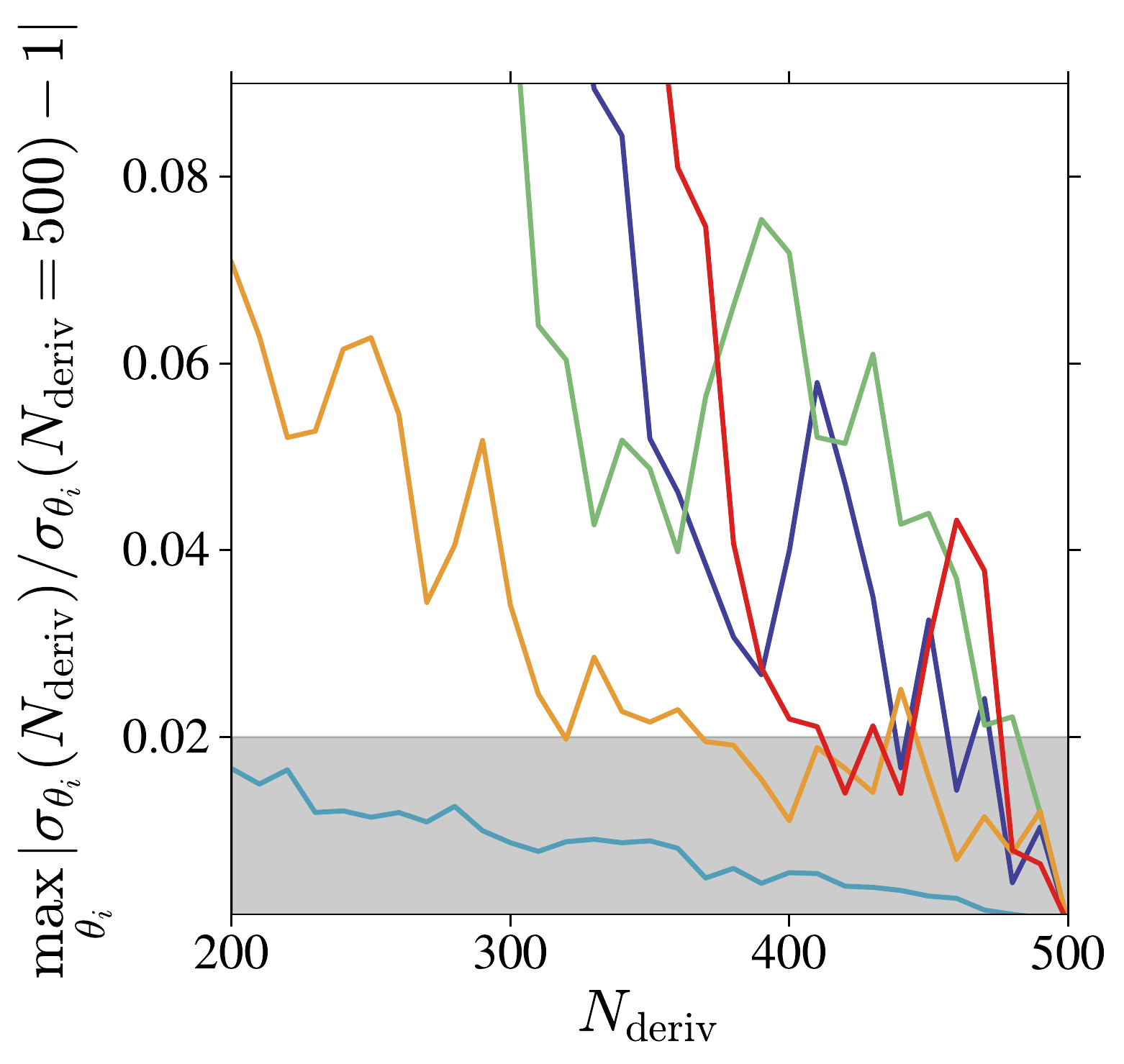}
        \caption{Convergence analysis of the numerical precision matrix as a function of $N_\mathrm{fid}$ with fixed $N_\mathrm{deriv}=500$ (upper panel) and derivatives as a function of $N_\mathrm{deriv}$ with fixed $N_\mathrm{fid}=7000$ (lower panel) for the combined spectra statistics $P_\mathrm{comb}$. The grey area shows the $2$\% agreement area.}
        \label{fig:convergence}
    \end{figure}

    In the Fisher forecast, we resort to numerical estimations of the precision matrices defined in Eq.~\eqref{eq:precision} but also of the derivatives from Eq.~\eqref{eq:derivatives} and \eqref{eq:derivatives_neutrinos}. To avoid biased results induced by a non-convergence of these quantities, it is essential to check the stability of the derived constraints under reduction of both $N_\mathrm{fid}$, the number of simulations used to compute the covariances and $N_\mathrm{deriv}$, the number of simulations for the derivatives. We focus here on the convergence of the constraints $\sigma_{\theta_i}$ derived in all setups, from individual environments with $40$ Fourier bins or from the combination of power spectra yielding the maximum total length among all the studied statistical summaries with $n=160$. Note that the convergence of the matter power spectrum in Quijote simulations is already studied in \cite{Villaescusa-Navarro2019}.
    In Fig.~\ref{fig:convergence}, we show how the maximum deviation of marginalised constraints behaves when varying $N_\mathrm{fid}$ in the upper panel and $N_\mathrm{deriv}$ in the lower one. We plot, for each considered statistic, the evolution of the maximum deviation over all cosmological parameters
    \begin{equation}
        \operatorname*{max}_{\theta_i} \left| \frac{\sigma_{\theta_i}(N)}{\sigma_{\theta_i}(N=N_\mathrm{max})} - 1 \right|,
    \end{equation}
    with $N$ being either $N_\mathrm{fid}$ or $N_\mathrm{deriv}$ and $N_\mathrm{max}$ respectively taking values $7000$ and $500$.
    For all the individual environments and their combination, convergence at a $\pm2$\% level is obtained when $N_\mathrm{fid} \sim 4500$ for the computation of the covariance matrix. Individually, the constraints from the several statistics reach a $5\%$ convergence when $N_\mathrm{deriv} \leq 400$ for all environments except walls for which the $5\%$ level is obtained around $450$ simulations. Convergence of the derivatives is however obtained at the percent level for $P_\mathrm{comb}$ when $N_\mathrm{deriv} \simeq 300$, highlighting the good convergence properties of this statistic and excluding any bias induced by numerical instabilities in the computation of Fisher constraints for the combination of spectra.
    
\section{Gaussianity of the statistics}

\begin{figure*}
    \centering
    \includegraphics[width=0.49\linewidth]{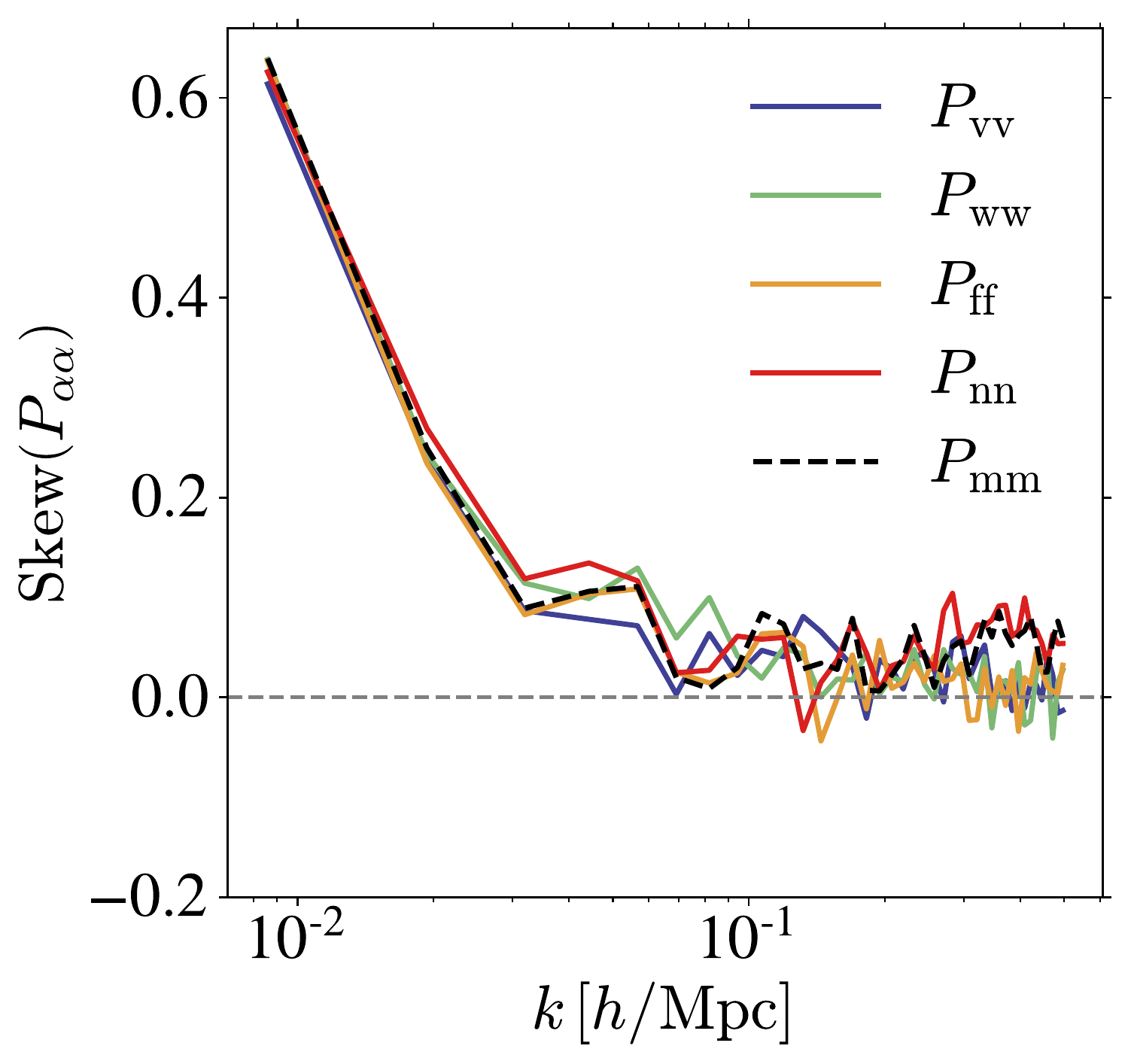}
    \includegraphics[width=0.49\linewidth]{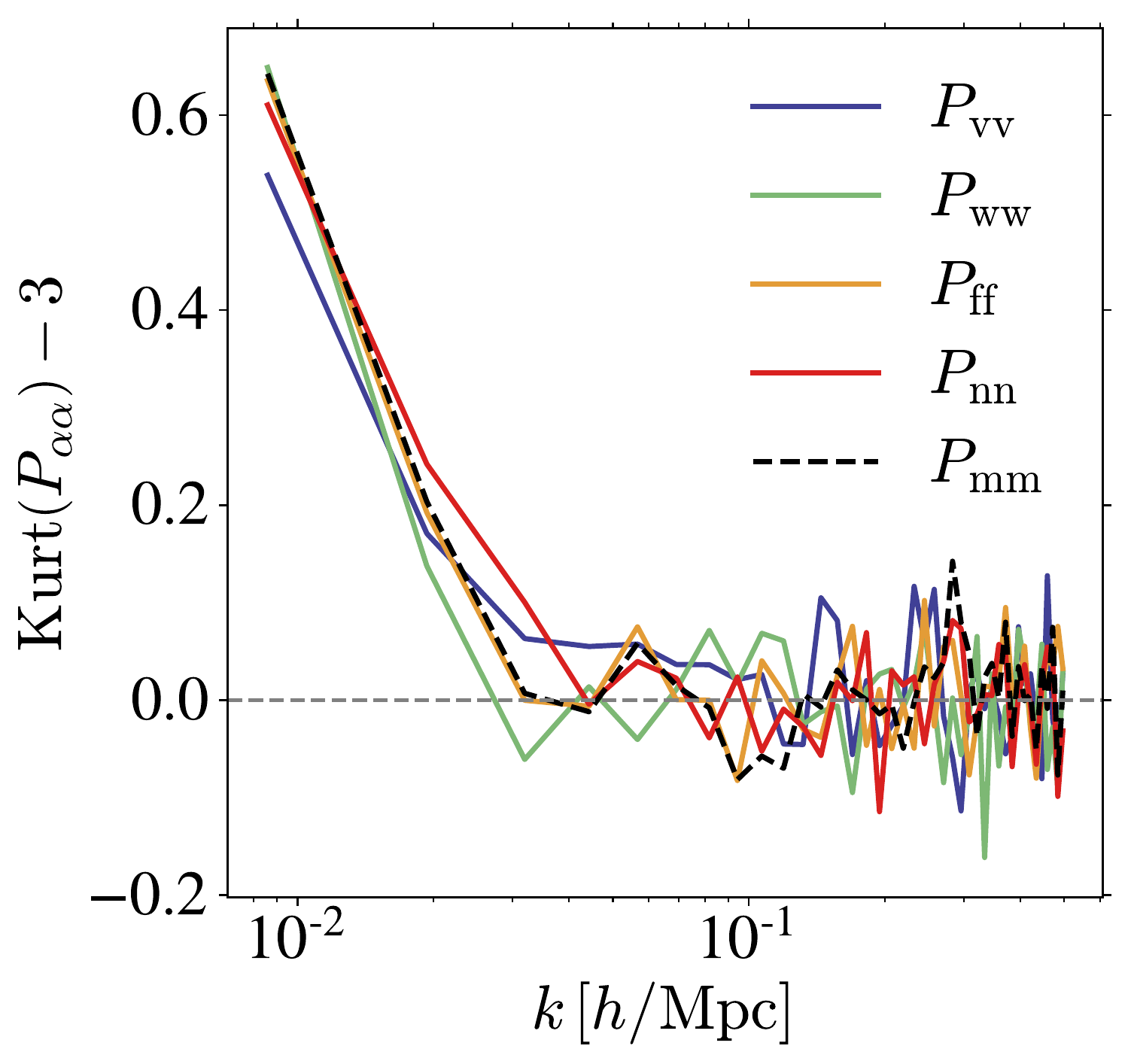}
    \caption{Skewness (left panel) and excess kurtosis (right panel) distributions for each $k$ bin of the power spectra. In both plots, the dashed grey horizontal line denotes the zero value.}
    \label{fig:gaussianity_real}
    \end{figure*}

    When deriving the Fisher matrices in Eq.~\eqref{eq:FIM_gaussian}, we assumed the distribution of the statistics $s$ to be Gaussian, expressed through Eq.~\eqref{eq:fisher_LL}. To quantify any deviation from Gaussianity which could induce biased estimation of errors, we report in Fig.~\ref{fig:gaussianity_real} the skewness and excess kurtosis distributions for each $k$ bin for all the computed power spectra. At large scales, the small number of modes available to compute the averages of Eq.~\eqref{eq:PS_real} creates significant non-Gaussianities for all spectra. As a result of the central limit theorem, at higher values of $k$, there are more modes for the estimators to be averaged over leading the Gaussian hypothesis to be more accurate. Still, some spectra, mainly $P_\mathrm{mm}$ and $P_\mathrm{nn}$, are showing deviations from the Gaussian hypothesis at small $k$ with a higher value of the skewness compared to other spectra. The impact of these non-Gaussian signatures are however limited on our analysis since we mostly focus on the gains of our statistics with respect to the matter power spectrum which already highlights the same features.
    
\section{Constraints with fixed segmentation parameters}

    In the main text, the presented results show the cosmological constraints obtained on $\Omega_\mathrm{m}$, $\Omega_\mathrm{b}$, $h$, $n_\mathrm{s}$, $\sigma_8$ and $M_\nu$ in Table \ref{tab:constraints_real} when marginalising over the two parameters of the T-web formalism used to identify the environments, namely $\lambda_\mathrm{th}$ and $\sigma_\mathcal{N}$. As expected, when fixing these at their fiducial values, respectively 0.3 and 2 $h$/Mpc, the constraints are tighter, as presented in Table \ref{tab:constraints_real_fixed}. This especially occurs for the $\sigma_8$ and $M_\nu$ parameters, while the others remain unchanged.

    \begin{table*}
    \centering
    \caption{Marginalised 1-$\sigma$ constraints obtained from the power spectrum in different environments for all cosmological parameters when fixing the nuisance parameters to their fiducial values, namely $\lambda_\mathrm{th} = 0.3$ and $\sigma_\mathcal{N} = 2$ Mpc/$h$. $\sigma_{M_\nu}$ is in unit of eV.}
    \label{tab:constraints_real_fixed}
    
    \renewcommand{\arraystretch}{1.5}
    \smallskip
    \begin{adjustbox}{max width=1.0\textwidth,center}
    \begin{tabular}{c|cccccc}
        \toprule
        Statistics & $\sigma_{\Omega_\mathrm{m}}$ & $\sigma_{\Omega_\mathrm{b}}$ & $\sigma_{h}$ & $\sigma_{n_\mathrm{s}}$ & $\sigma_{\sigma_\mathrm{8}}$ & $\sigma_{M_\nu}$ \\
        \midrule
        $P_\mathrm{mm}$ & $0.0969$ & $0.0413$ & $0.5145$ & $0.5019$ & $0.0132$ & $0.8749$ \\
        $\color{void} P_\mathrm{vv}$ & $0.0305 \, (3.2)$ & $0.0183 \, (2.3)$ & $0.1907 \, (2.7)$ & $0.1104 \, (4.5)$ & $0.0110 \, (1.2)$ & $0.3272 \, (2.7)$ \\
        $\color{wall} P_\mathrm{ww}$ & $0.0340 \, (2.9)$ & $0.0204 \, (2.0)$ & $0.1884 \, (2.7)$ & $0.0688 \, (7.3)$ & $0.0250 \, (0.5)$ & $0.5183 \, (1.7)$ \\
        $\color{filament} P_\mathrm{ff}$ & $0.0174 \, (5.6)$ & $0.0124 \, (3.3)$ & $0.1200 \, (4.3)$ & $0.0684 \, (7.3)$ & $0.0125 \, (1.1)$ & $0.2708 \, (3.2)$ \\
        $\color{node} P_\mathrm{nn}$ & $0.0271 \, (3.6) $ & $0.0191 \, (2.2) $ & $0.2013 \, (2.6)$ & $0.1350 \, (3.7)$ & $0.0242 \, (0.5)$ & $0.5365 \, (1.6)$ \\
        
        $\color{combination} P_\mathrm{comb}$ & $0.0123 \, (\bm{7.9})$ & $0.0092 \, (\bm{4.5})$ & $0.0782 \, (\bm{6.6})$ & $0.0307 \, (\bm{16.4})$ & $0.0018 \, (\bm{7.2})$ & $0.0360 \, (\bm{24.3})$ \\
        \bottomrule
    \end{tabular}
    \end{adjustbox}
    \end{table*}
    
\end{appendix}

\end{document}